\documentclass[a4paper,11pt]{article}
\usepackage{jinstpub} 
\usepackage{lineno,hyperref}
\usepackage{gensymb}
\usepackage{subcaption}
\usepackage{graphicx,tabularx}
\usepackage{ifthen}
\newcolumntype{P}[1]{>{\centering\arraybackslash}p{#1}}

\title{\boldmath Temperature dependence of the long-term annealing behavior of neutron irradiated diodes from 8-inch p-type silicon wafers}

\author[a,1]{Leena Diehl\note{Corresponding author.}}
\author[a,b]{Oliwia~Agnieszka~Ka\l{}uzi\'{n}ska}
\author[a,c]{Marie~Christin~M\"uhlnikel}
\author[a,d]{Max Andersson}
\author[a,e]{Natalya Gerassyova}
\author[a,f]{Jenan Amer}
\author[a]{Eva~Sicking}
\author[a,g]{Dana Groner}
\author[a,b]{Jan Kieseler}
\author[a]{Matteo~Defranchis}

\affiliation[a]{Experimental Physics department, CERN,\\Espl. des Particules 1, Geneva, Switzerland}
\affiliation[b]{Karlsruhe Institute of Technology,\\Engelbert-Arnold-Straße 4, Karlsruhe, Germany}
\affiliation[c]{Kirchhoff Institute for Physics, Heidelberg University,\\Im Neuenheimer Feld 227, Heidelberg, Germany}
\affiliation[d]{Uppsala University, Box 35, 751 03 Uppsala, Sweden}
\affiliation[e]{Al-Farabi Kazakh National University \\Al-Farabi Avenue 71, Almaty 050040, Kazakhstan}
\affiliation[f]{An Najah National University,\\ Old Campus Street 7, Nablus,Palestine}

\affiliation[g]{HAN University of Applied Sciences,\\Ruitenberglaan 29, 6826 CC Arnhem, Netherlands}
\emailAdd{leena.diehl@cern.ch}

\abstract{
To face the higher levels of radiation due to the 10-fold increase in integrated luminosity during the High-Luminosity LHC, the CMS detector will replace the current  Calorimeter Endcap (CE) using the High-Granularity Calorimeter (HGCAL) concept. The high-radiation regions of the the CE, where fluences between $\rm1.0\cdot10^{14}~n_{eq}/cm^{2}$ and  $\rm1.0\cdot10^{16}~n_{eq}/cm^{2}$ and doses of up to 2\,MGy are expected considering an integrated luminosity of $3\,\rm ab^{-1}$, will be equipped with silicon pad sensors. This includes the entire electromagnetic as well as parts of the hadronic section of the CE. 

The silicon sensors are processed on 8-inch p-type wafers with an active thickness of 300\,\textmu m, 200\,\textmu m and 120\,\textmu m and cut into hexagonal shapes for optimal use of the full wafer area and tiling. With each main sensor, several small test structures (e.g. pad diodes) are hosted on the wafers, used for quality assurance and radiation hardness tests. In order to investigate the radiation-induced bulk damage, these diodes have been irradiated with reactor neutrons at JSI (Jožef Stefan Institute, Ljubljana, Slovenia) to fluences between $\rm5\cdot10^{14}~n_{eq}/cm^{2}$ and $\rm1.5\cdot10^{16}~n_{eq}/cm^{2}$.

This study focuses on the isothermal annealing behavior of the bulk material at different temperatures between $5.5^\circ$C and $60^\circ$C using electrical characterization and charge collection measurements. The results are used to extract the annealing time constants for this material and fluence range based on the Hamburg model approach to allow an estimation of the expected annealing effects in silicon sensors during the year-end technical stops and the long HL-LHC shutdowns. The annealing parameters found in this study will make it possible to model the annealing behavior of p-type silicon detector projects at HL-LHC fluence ranges better than the existing Hamburg model. }

\keywords{Radiation damage, silicon sensors, annealing, Hamburg model}

\begin{document}
\maketitle
\flushbottom
\section{Introduction}
\label{sec:intro}

The High-Luminosity Large Hadron Collider (HL-LHC) upgrade aims at enhancing the performance of the LHC, the most powerful particle accelerator in the world, to boost the potential for scientific discoveries starting from 2030.
The integrated luminosity will be increased by a factor of ten, which poses significant challenges in terms of radiation tolerance and event pileup for the detectors~\cite{Apollinari:2284929}.

As part of the HL-LHC upgrade within the CMS Experiment~\cite{CMS_det_paper}, the current Calorimeter Endcap (CE) will be replaced, using the novel High Granularity Calorimeter (HGCAL) concept \cite{HGCAL-TDR}.
Silicon sensors were selected as active material for the majority of the CE upgrade due to their compactness, rapid signal formation, and adequate radiation hardness.
The silicon sensors are fabricated on 8-inch wafers and diced to form a hexagonal shape for efficient use of the full wafer area and tiling~\cite{ HGCAL-TDR,CMSHGCAL:2022mtx}.

In this measurement campaign, the isothermal annealing behavior of the silicon bulk material was investigated using dedicated 5x5 $\rm mm^2$ test structure diodes from the edge region of the wafer, irradiated with reactor neutrons up to a fluence of $\rm1.5\cdot10^{16}$ 1-MeV neutron equivalents per square centimeter $(\rm n_{eq}/cm^{2})$, accounting for a $+50\%$ safety factor above the expected HL-LHC levels. 35 diodes were irradiated to five different fluences and split into five different batches that were long-term annealed at different temperatures in the range of $5.5^\circ$C to $60.0^\circ$C, since $60^\circ$C is the standard annealing temperature for silicon sensors, and  $5.5^\circ$C was the closest achievable temperature to the currently foreseen $0^\circ$C shutdown temperature of HGCAL with the available infrastructure. The study aims to evaluate the temperature dependence of annealing in order to extrapolate the amount of annealing during the foreseen shutdown periods of the HL-LHC and to verify that the performance of the silicon sensors will meet the requirements until the end. 
Through leakage current (IV), capacitance (CV) and Transient-Current-Technique (TCT, \cite{kramberger2014advanced}) measurements, the key parameters such as the current-related damage rate, saturation voltage and charge collection were measured and used as input for the so-called Hamburg model~\cite{Moll:2018fol,Moll:1999kv} to extract the annealing parameters.

\section{Devices under test}
\label{sec:sensors}

The silicon sensors to be used in the CMS Calorimeter Endcap upgrade~\cite{HGCAL-TDR} consist of DC-coupled, planar, high resistivity ($>$3 k$\Omega$cm), p-type hexagonal silicon sensors with a crystal orientation of $<100>$ produced on 8-inch circular wafers by Hamamatsu Photonics K.K\footnote{\url{https://www.hamamatsu.com/eu/en.html}}.
Sensors are produced in three active thicknesses and two production processes. The varying thicknesses are to adapt to the differences in the expected fluence in HGCAL, where the thicker sensors will be used in lower fluence regions to benefit from an initially larger signal, while the more radiation hard thin sensors will equip the high fluence regions.
The wafers holding the 300~\textmu m and 200~\textmu m sensors are produced in the float zone (FZ) process, thinning down a 600~\textmu m thick wafer, while the wafers holding the 120~\textmu m sensors are produced in the epitaxial (EPI) process on top of a low-resistivity handling wafer thinned to $\sim$180 \textmu m thickness. The epitaxial process was chosen for the  120~\textmu m sensors  as they would be too fragile without an additional handling wafer.
The remaining space on the wafer is used for the fabrication of small-sized test structures \cite{dissPQC}, including the single square diodes with a side length of 5~mm used for this study. 

The samples were irradiated with neutrons to fluences between $\rm2\cdot10^{15}\,n_{eq}/cm^{2}$ and $\rm1.5\cdot10 ^{16}\,n_{eq}/cm^{2}$ in the TRIGA reactor located at the Jožef Stefan Institute in Ljubljana, Slovenia~\cite{SNOJ2012483, AMBROZIC2017140}. The fluence uncertainty for all samples was estimated at $10\%$ from the irradiation facility in Ljubljana. The samples were separated into five different batches that were annealed at the five annealing temperatures of $5.5^\circ$C, $20.5^\circ$C,  $30.0^\circ$C,  $40.0^\circ$C and  $60.0^\circ$C. All batches contain 7 sensors with the fluences and thicknesses summarized in Table~\ref{tab:Overview}. The batch annealed at $40^\circ$C has only six sensors, not including a 200\,\textmu m sensor with a fluence of $\rm8\cdot10^{15}\,n_{eq}/cm^{2}$ because 2 of the 6 sensors sent for irradiation were broken during transport and no sample was available any more. 

\begin{table}
    \centering
        \caption{Overview of the samples per batch including fluence, thickness and corresponding initially calculated and revised time of annealing at $60^\circ$C with equivalent effect as annealing during irradiation, separated for floatzone ($200\,\mu$m and $300\,\mu$m) and epitaxial ($120\,\mu$m) sensors as discussed in Section \ref{sec:Neff}. For the initial estimation the Hamburg model values, which are material independent, were used for both epitaxial and floatzone sensors.}
    \begin{tabular}{|P{21mm}|P{21mm}|P{21mm}|P{30mm}|P{30mm}|}
    \hline
    Fluence [$\mathrm{n_{eq}\,cm^{-2}}$] & Irradiation time [min] & Thickness [$\mu$m]
 & Est. initial $60^\circ$C \hspace{0.2cm}[min] & Corr. est. initial $60^\circ$C [min]\\
         \hline
          $2.0\cdot10^{15}$ & 21.8 & 300 & $1.8\substack{+1.7 \\ -0.9}$& $2.0\substack{+1.7 \\ -1.0}$\\\hline
         $4.0\cdot10^{15}$ & 43.3 & \hspace{0.25cm} 200 \newline 300 & \vspace{0.01cm}$7.7\substack{+6.3 \\ -3.6}$ &\vspace{0.01cm} $7.8\substack{+6.4 \\ -3.6}$\\\hline
        $6.0\cdot10^{15}$ & 64.9 &\hspace{0.25cm}  120 \newline 200 & \vspace{0.01cm}$14.0\substack{+ 11.5\\ -6.5}$ & \hspace{0.6cm}$14.2\substack{+ 11.6\\ -6.5}$ \newline $16.2\substack{+11.5 \\-6.9}$ \\\hline
          $8.0\cdot10^{15}$ & 86.8 & 200 & $20.3\substack{+16.6 \\ -9.4}$ &$20.7\substack{+ 16.7\\ -9.4}$\\\hline
          $1.5\cdot10^{16}$ & 161.8 & 120 & $39.9\substack{+32.8 \\ -18.4}$ & $ 45.9\substack{+ 32.5\\ -19.4}$\\
        
         \hline
             \end{tabular}

    \label{tab:Overview}
\end{table}

Each sample is placed on a custom-designed 2-layer PCB with electrically conductive silver paint and wire-bonded to gold contact pads of SMA readout connectors. The high voltage is applied to the diode’s backside, while the n+ implant is grounded through the readout circuit. The implant is surrounded by a guard ring, which is grounded during the measurements.  In order to control the temperature of the diode, in the following referred to as device under test (DUT), during measurements and annealing steps, a PT1000 resistor is glued close to the DUT. 

Between measurements, the samples are annealed in either a dedicated oven on pre-heated copper blocks for $(30\pm0.5)^\circ$C, $(40\pm0.5)^\circ$C or $(60\pm0.5)^\circ$C, in a temperature stabilized cleanroom at an average temperature of $(20.5\pm1)^\circ$C or a dedicated refrigerator running at an average temperature of $(5.5\pm1.5)^\circ$C. Temperatures are monitored through dedicated temperature and humidity sensors placed in the cleanroom and the fridge. For the oven annealing, it was recorded with a logger using the PT1000 glued next to the sensors for one of the placed samples.

During irradiation, the sensors are not cooled and can warm up from the steady state around $20-25^\circ$C to temperatures of around $50^\circ$C \cite{cindro2019measurement}. An uncertainty of $5^\circ$C is assumed for the maximum temperature. Therefore, non-negligible annealing occurs already during irradiation, which is accounted for by calculating the equivalent annealing time at a certain temperature, assuming all annealing happened after the full fluence was delivered. 
For the initial analysis, this was done using the parameters of the Hamburg model \cite{Moll:2018fol}. After recording all data and extracting the same parameters for the currently used p-type sensors and fluence range of this study (described in detail in Section \ref{sec:Neff}), the calculation of the in-reactor annealing was re-evaluated to correct the offset, which resulted in a shift especially for the low annealing temperatures.  A comparison of the initial annealing times at $60^\circ$C annealing obtained with the established and revised parameters is shown in Table ~\ref{tab:Overview}. The results presented in the following sections are using the corrected offsets for all annealing temperatures.
Between measurements or annealing steps, the sensors are kept in a freezer at $-18^\circ$C to limit  uncontrolled annealing.

\section{Experimental Setup and Measurements}
\label{sec:setup}

\begin{figure}[h]
    \centering
    \includegraphics[width=0.8\linewidth]{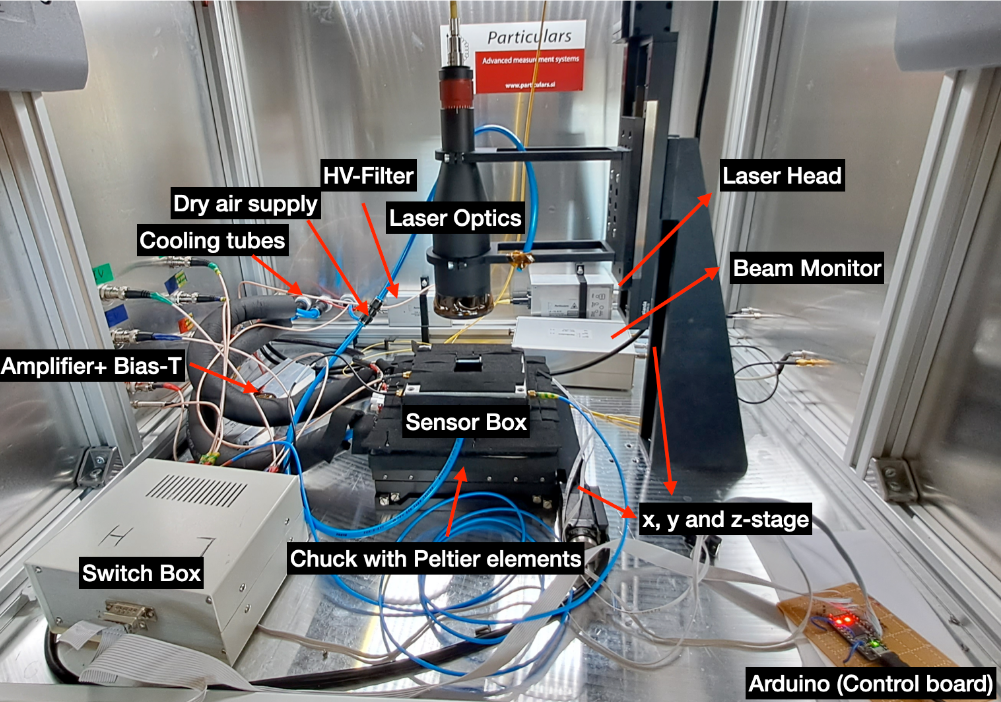}
    \caption{IV-CV-TCT setup used to obtain all data presented, not showing the measurement instruments outside of the Faraday cage \cite{kaluzinska2025annealing}.}
    \label{fig:Setup}
\end{figure}

All measurements are taken in an IV-CV-TCT setup based on a laser-system provided by Particulars\footnote{\url{https://particulars.si/index.php}}, shown in Figure~\ref{fig:Setup}.
The PCBs with the wirebonded sensors are placed on a copper holder inside a sensor box and connected to the readout. Through a switchbox, the different circuits for the individual measurements are connected (schematics in \cite{kaluzinska2025annealing}). 
For all measurements, the sensors are measured at a temperature of $-20^\circ$C, achieved through a combination of Peltier-elements and a chiller set pumping ethanol with a temperature of  $-30^\circ$C through the chuck below the mounting stage under the sensor box. The systematic uncertainties of all measurements have the common contribution of the sensor temperature uncertainty in the setup, which is assumed to be $\pm1^\circ$C.

The leakage current is measured with a picoammeter in the HV-line while readout pad and guard ring are grounded. The uncertainty due to the temperature is calculated by scaling the current to $-19^\circ$C and $-21^\circ$C and substracting the measured value at $-20^\circ$C. The recorded uncertainties of the picoammeter during the measurements are negligible in comparison.

The capacitance is measured in parallel mode at a frequency of 2\,kHz, using a low voltage amplitude of 0.5\,V. 
In unirradiated sensors, the measured capacitance of a sensor stays constant for voltages beyond the depletion voltage, giving the opportunity to evaluate the depletion voltage by extracting the intersection of two linear fits on the $1/C^2$ data, one for the rising part and one for the constant part. 
However, for irradiated sensors the concept of depletion is no longer well defined, which is why this intersection voltage is referred to as the "saturation voltage" in this work. Two important factors in this are the temperature and frequency dependencies of the CV measurements of irradiated silicon sensors \cite{Kieseler_2023, CAMPBELL2002402}. The temperature uncertainty is taken into account in the measurement uncertainties, while the frequency is the same for all measurements to keep it consistent through the annealing. While the frequency dependence varies for different fluences, it was found that the annealing behavior of the extracted saturation voltages does not change with frequency \cite{Kieseler_2023}. 

The extraction of saturation voltages is only feasible for some sensors in this study, as especially for the $300\,\upmu$m sensors the saturation voltages exceed the measurement limit of 900\,V significantly. 
For thinner sensors at lower fluences the extraction is possible for short annealing times, while the saturation voltages start exceeding 900\,V beyond a certain annealing time, which is dependent on the fluence (see Figure \ref{fig:CV-curves}). As it was observed that the capacitance beyond depletion remains constant for all sensors of each thickness, independent of fluence or annealing time, for some measurements saturation voltages slightly above 900\,V could be extracted by only fitting the rising part of the data and calculating the intersection with the constant capacitance measured in the sensors in previous measurements.

Infrared laser light of 1064\,nm wavelength with a frequency of 1\,kHz is injected from the front side of the DUT (IR-top TCT) directed at a 1\,mm-diameter hole in the center of the metal contact layer of the diode. The laser intensity is tuned to the charge that 40~MIPs would deposit in $300\,\mu$m silicon, well within the linear region of the laser and the diode. It is monitored daily using the same unirradiated 300\,\textmu m sensor used for the calibration to apply correction factors accordingly to the measured collected charge if necessary. 
This is to account for an uncertainty regarding the laser stability, besides the temperature uncertainty. Environmental temperature variations slightly influence the laser intensity, as well as the temperature stability of the DUT. Therefore, a clear distinction between the influences due to the temperature-dependent laser absorption in silicon and the laser intensity is not possible. Hence, to account for this, an overall charge collection uncertainty of 5.0\% was estimated based on the observed fluctuations during the two year measurement period.

For each voltage, 200 averaged waveforms are recorded, with a single waveform averaging 50 samples. The collected charge is calculated by integrating over the signal waveform and dividing by the amplifier gain. Exemplary waveforms can be found in \cite{kaluzinska2025annealing}.

 The sensor box is mounted on an xy-stage to position the DUT, while the laser is moved by a separate z-stage in order to find the focus point. Dry air is flushed into the Faraday cage and into the sensor box to reduce the risk of condensation and frost.
 A photodiode as a beam monitor is included as a reference for the laser intensity, using a fiber splitter to direct 50\% to the beam monitor, while 50\% goes through the optics system to the DUT. 
 A wide-band amplifier with 53 dB gain by the Particulars company is used for the DUT signal, with a bias-T with a $50\,\Omega$-terminated HV-line placed before it to reduce noise and to protect the amplifier. Additionally, a HV-filter is used to remove additional noise from the HV-line, where a second bias-T with a $50\,\Omega$-terminated signal line was implemented as well to reduce a specific reoccurring periodic noise halfway through the campaign. The sensor is still biased from the backside as for IV and CV measurements. This did not have a noticeable impact on the collected charge measurements. 
 More details about the setup can be found in \cite{kaluzinska2025annealing}. 


\section{Fluence dependence}
\label{sec:Fluence_dep}
\begin{figure}
    \centering
    \begin{subfigure}{0.48\textwidth}
        \includegraphics[width=\textwidth]{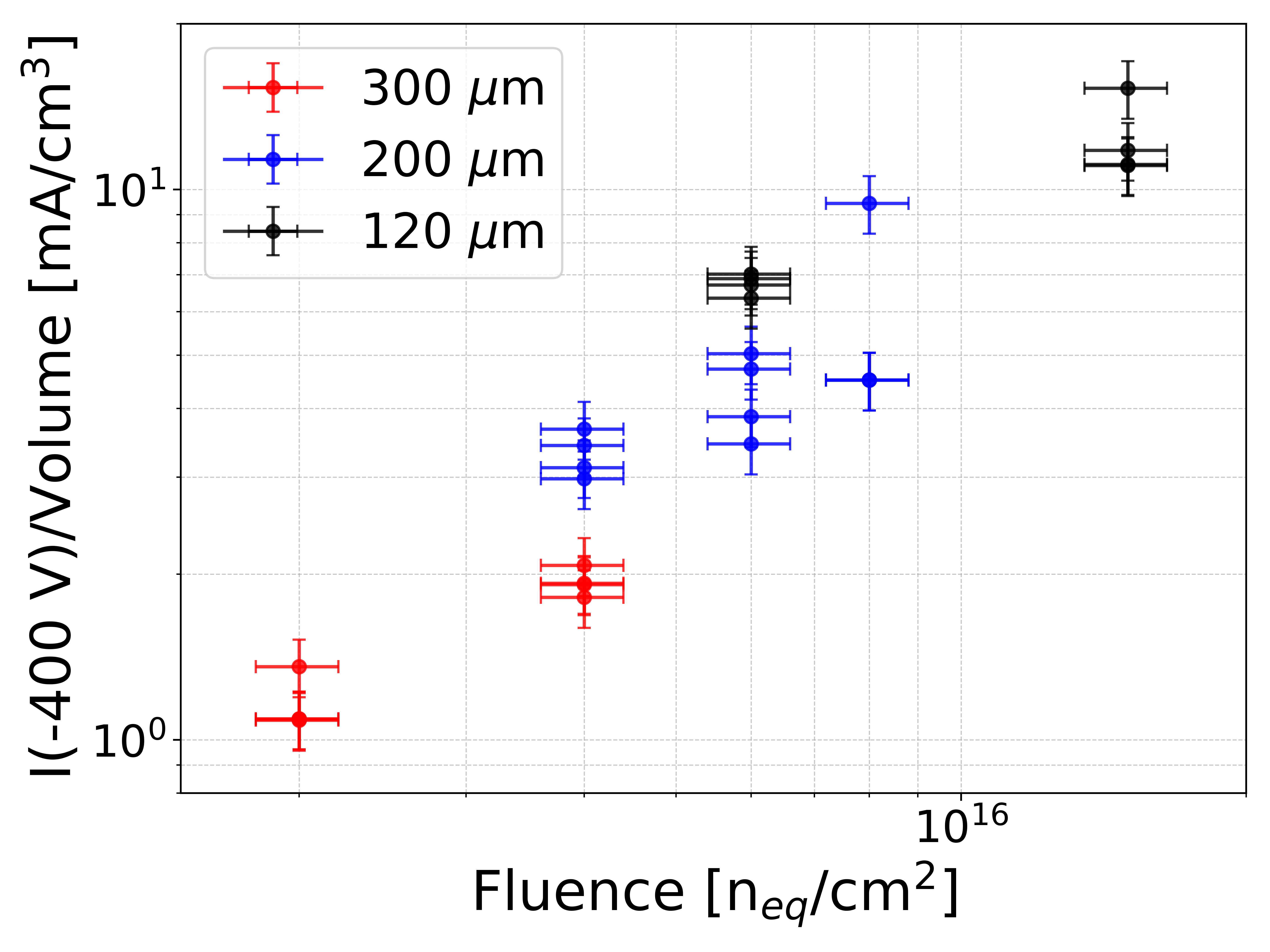}
        \caption{Volume normalized leakage current at $400\,$V as a function of fluence.}
        \label{fig:Leak_vs Fluence}
    \end{subfigure}
    \hfill
    \begin{subfigure}{0.48\textwidth}
        \includegraphics[width=\textwidth]{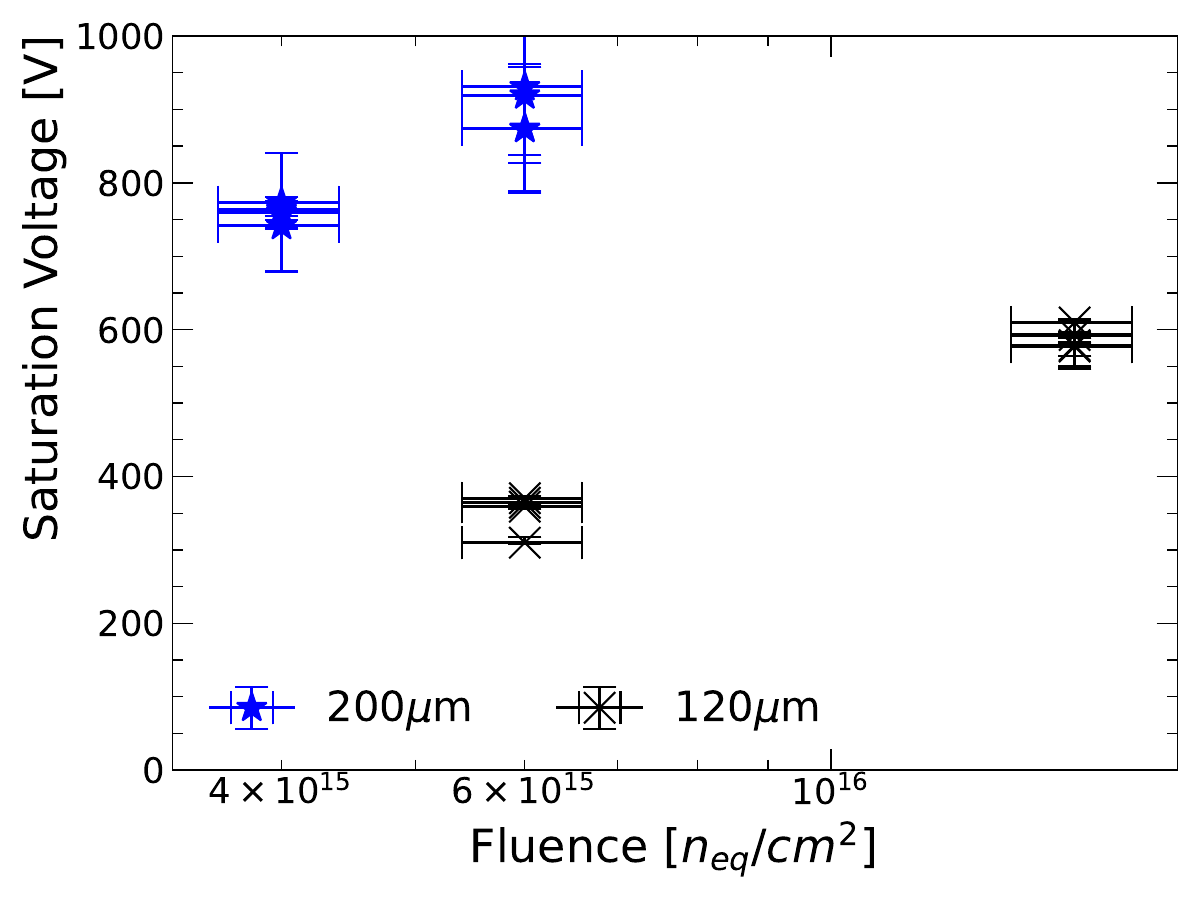}
        \caption{Saturation voltage as a function of fluence.}
       \vspace{0.45cm}
        \label{fig:Sat_Volt}
    \end{subfigure}
\caption{Volume normalized leakage current at 400\,V and extracted saturation voltages of all sensors without any additional annealing as a function of fluence.}
\label{Fig:Fluence_dep}
\end{figure}

Before any additional annealing was performed, all sensors have been measured and the results were evaluated as a function of fluence. 
Figure~\ref{Fig:Fluence_dep} presents the leakage current at $-20^\circ$C at 400\,V (a) and the saturation voltage (b) in dependence of the fluence.

As expected, the leakage current increases linearly with fluence for all thicknesses even up to the highest fluence of $1.5\cdot10^{16}\,\rm n_{eq}/cm^{2}$. Despite the volume normalization, a small thickness dependence seems to be present in the leakage current, as thinner sensors exhibit a relatively higher current at the same fluences. This is influenced by the usage of a fixed voltage instead of taking the leakage current at the depletion voltage, which is not possible given the high fluences.
400\,V was chosen as it is low enough that no additional high electric field effects as e.g. avalanche multiplication play a role, especially with long-term annealing.

For the saturation voltage, not all sensors could be evaluated because  the saturation voltage exceeded the measurement limit of 900\,V significantly. However, the expected increase with fluence is still visible for both floatzone and epitaxial sensors.
Additionally, the thicker samples exhibit a significantly larger saturation voltage at the same fluence as expected. This plot confirms that at 400\,V electric field is not established across the entire sensor thickness as the saturation voltage is significantly higher  for most of them, affecting the leakage current at this voltage.

\section{Annealing of Leakage Current}
\label{sec:Ileak}
\begin{figure}
    \centering
    \begin{subfigure}{0.49\textwidth}
        \includegraphics[width=\textwidth]{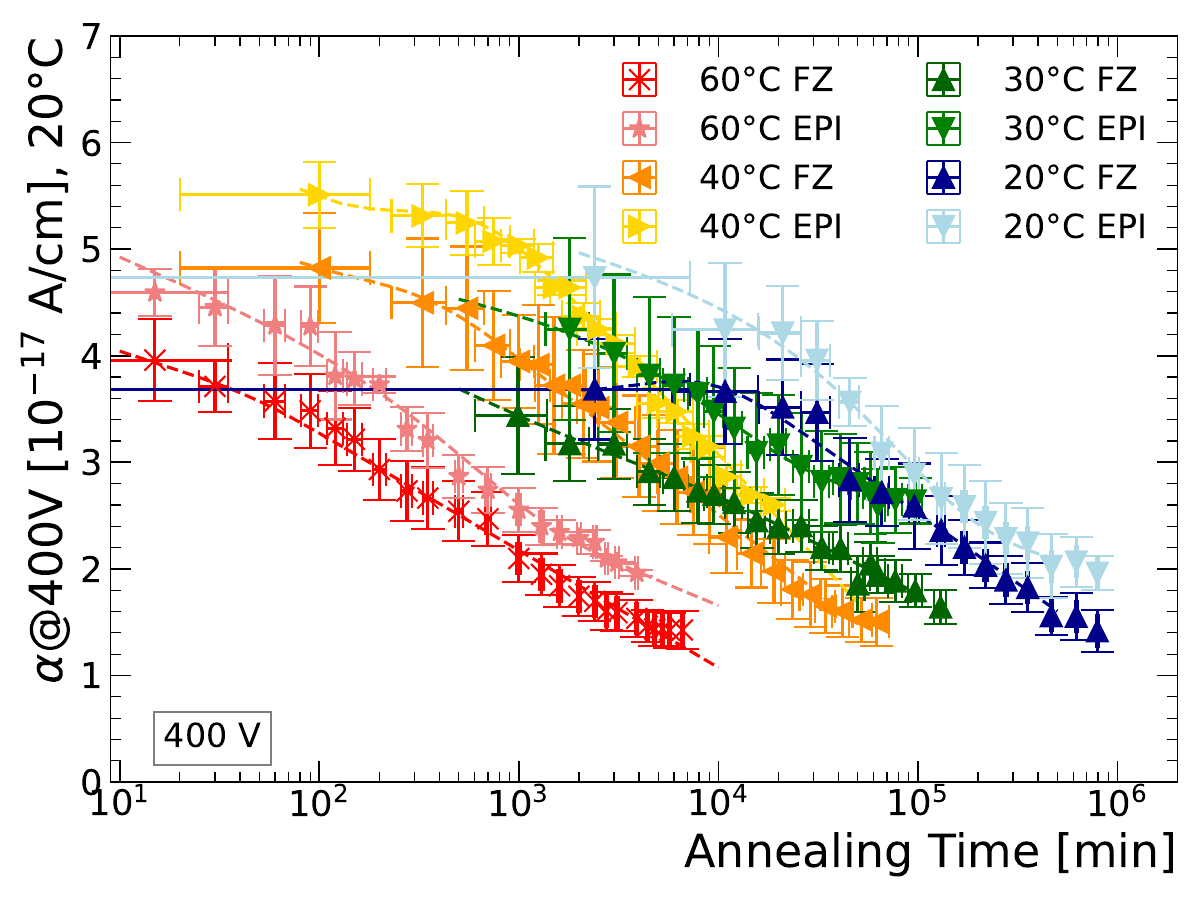}
        \caption{Damage parameter $\alpha$ in dependence of annealing time for each annealing temperature.}
        \label{fig:alpha_common}
    \end{subfigure}
    \begin{subfigure}{0.48\textwidth}
        \includegraphics[width=0.97\textwidth]{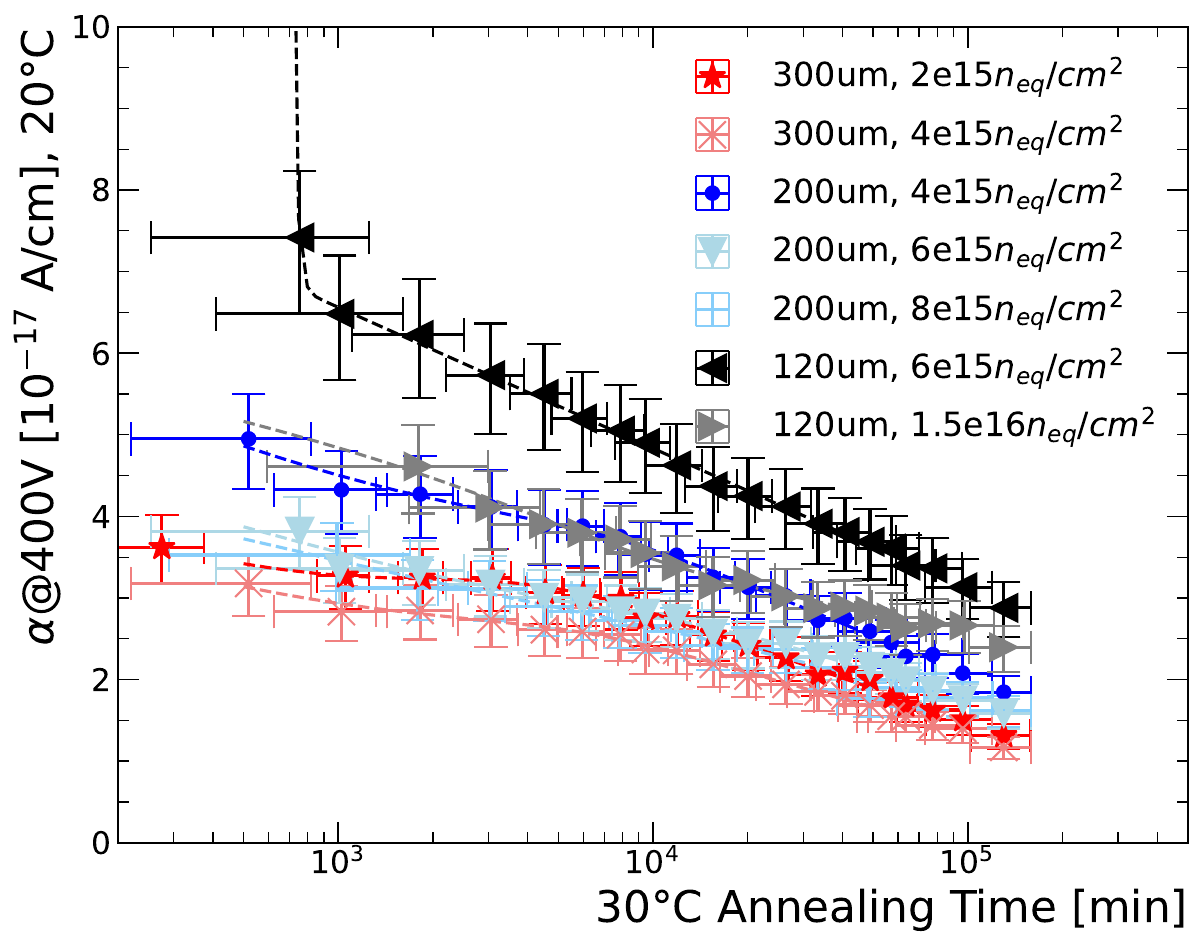}
        \caption{Damage parameter $\alpha$ in dependence of annealing time at $30^\circ$C for each sensor.}
        \label{fig:alpha_individual}
    \end{subfigure}
\caption{Damage parameter extracted from leakage current measurements at 400\,V, the dashed lines represent the Hamburg model fits.}
\label{Fig:Leakage}
\end{figure}

The leakage current was measured for all sensors after each annealing step as a function of the bias voltage in steps of 25\,V from 0\,V to 900\,V. The measured leakage current $I$ at 400\,V is scaled to $20^\circ$C using temperature scaling of the reverse current and used to calculate the current related damage rate $\alpha$. This can be done either for each sensor individually, by dividing the volume normalized current by the fluence $\Phi$, or for each annealing temperature for all sensors. This is done by plotting the volume $V$ normalized current as a function of fluence and fitting the equation
\begin{equation}
 I/V=\alpha \Phi
\end{equation}
to the data. This is done for each annealing step. For $30^\circ$C,$40^\circ$C and $60^\circ$C the sensors were brought to a similar annealing time within the first annealing steps by annealing the sensors individually so that they all were at the same total annealing times afterwards. For the $20^\circ$C annealing, the different in-reactor annealing times calculated using the Hamburg model were significantly longer and no individual annealing was done. To still calculate the $\alpha$-values for a certain annealing time, the data was grouped together according to the total annealing time and the average between different annealing times was taken as annealing time for this $\alpha$ value. 
The damage parameter $\alpha$ is then plotted as a function of annealing time. According to the Hamburg model \cite{Moll:1999kv}, the dependence on annealing time $t$ can be described with 
\begin{equation}
    \alpha (t) = \alpha_I \exp{(-t/\tau_I)} + \alpha_0-\beta\ln{(t/t_0)}
    \label{Ileakeq}
\end{equation}
where $\alpha_I$ is the amplitude and  $t_0$ is set to 1 minute. The first term describes the short-term annealing, and the second term the long-term annealing. 
The annealing time constants $\tau_I$ represent the temperature dependence of the leakage current annealing. 
The damage parameter in dependence of the annealing time for the different annealing temperatures is presented in Figure~\ref{fig:alpha_common}. Due to the longer equivalent in-reactor annealing times at lower temperatures the $\alpha$ data is starting at later annealing times.  As can be seen, the exponential decrease expected for the short-term annealing is not visible, whereas the logarithmic long-term decrease follows the expectation. However, the Hamburg model fits include the exponential decrease in the short time range. Thus, the lack thereof in the measured data leads to fits with equation \ref{Ileakeq} where the parameters do no longer have a physical meaning, such as negative amplitudes $\alpha_I$ or annealing time constants $\tau_I$ that show no temperature dependence or even exceed the recorded annealing time. This could be caused by several factors or a combination of them. First, the leakage current is evaluated at constant voltage instead of the depletion voltage, leading to differences in how much of the sensor volume has a high electric field, significantly influencing the leakage current. Secondly, the long in-reactor annealing times for the high-fluence sensors, as well as the limited number of measurement steps available in the short-term annealing time frame. Furthermore, due to the high fluences, the recorded current can be influenced by other effects, such as the formation of a double-junction \cite{VERBITSKAYA2006528, Agram:23731}, related to the interaction of traps with the leakage current, which has a temperature and fluence dependence. Additionally, the total current was recorded instead of the pad current only. Therefore, the sensor area is less well defined in comparison to using only the pad current, and the current of the guard ring is contributing and potentially having significant effects during the annealing. This was considered for follow-up studies, where only the pad current is measured. 
Thus, annealing parameters cannot be extracted from the leakage current data with the current model. The same was observed when looking at each sensor individually, as shown exemplary in Figure~\ref{fig:alpha_individual} for the sensors annealed at $30^\circ$C.

Additionally, while the $\alpha$ values are not fully in agreement for all sensors, there is no clear fluence dependence as expected. The extracted values seem to vary for different thicknesses despite the volume normalization, with higher values for lower thicknesses. This is again influenced by the usage of a fixed voltage instead of taking the leakage current at the depletion voltage.

\begin{figure}
    \centering
    \begin{subfigure}{0.48\textwidth}
        \includegraphics[width=\textwidth]{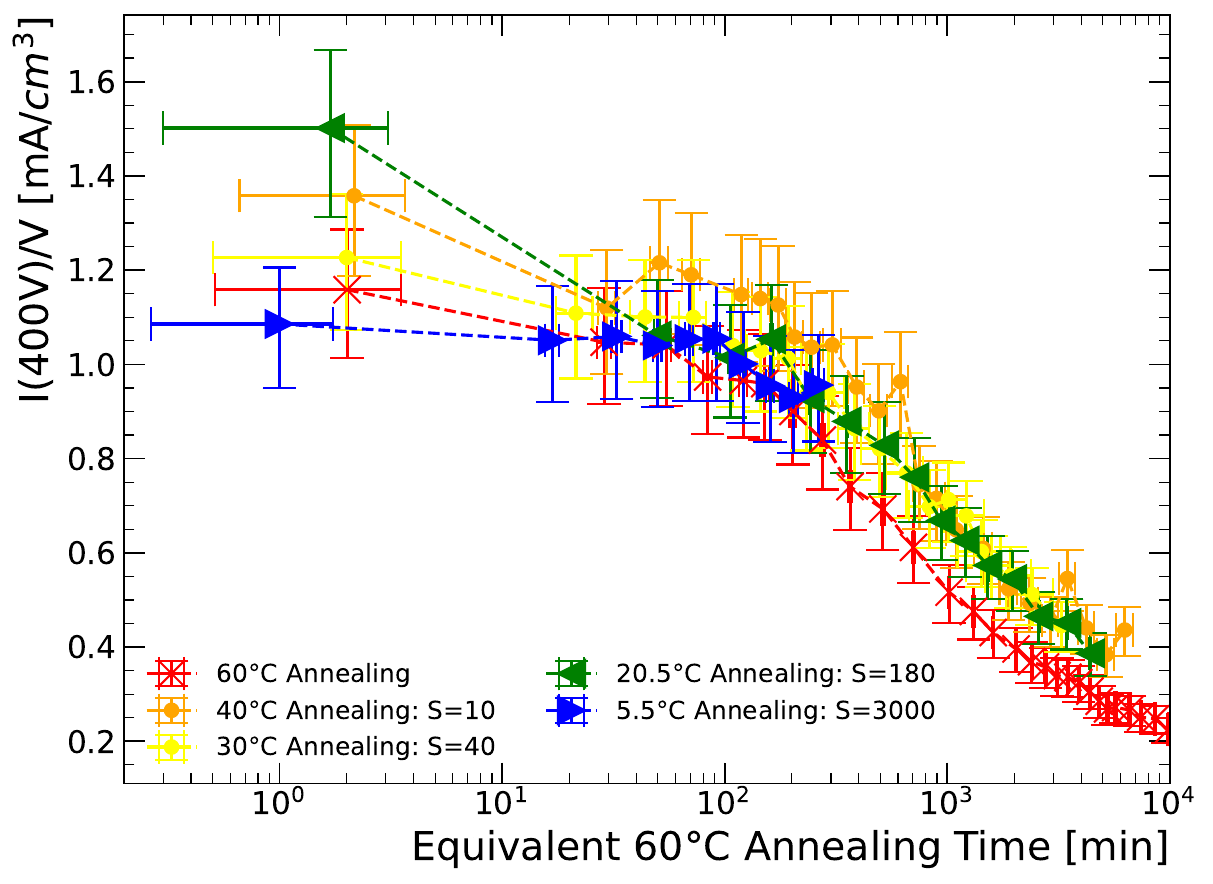}
        \caption{Volume normalised leakage current of the $300\upmu$m sensors irradiated to $2\cdot10^{15}\,\rm n_{eq}/cm^{2}$ at 400\,V for different annealing temperatures scaled to the $60^\circ$C annealing time. The lines are shown as eye-guide.}
        \label{fig:leak_timescaled}
    \end{subfigure}
    \hfill
    \begin{subfigure}{0.48\textwidth}
        \includegraphics[width=\textwidth]{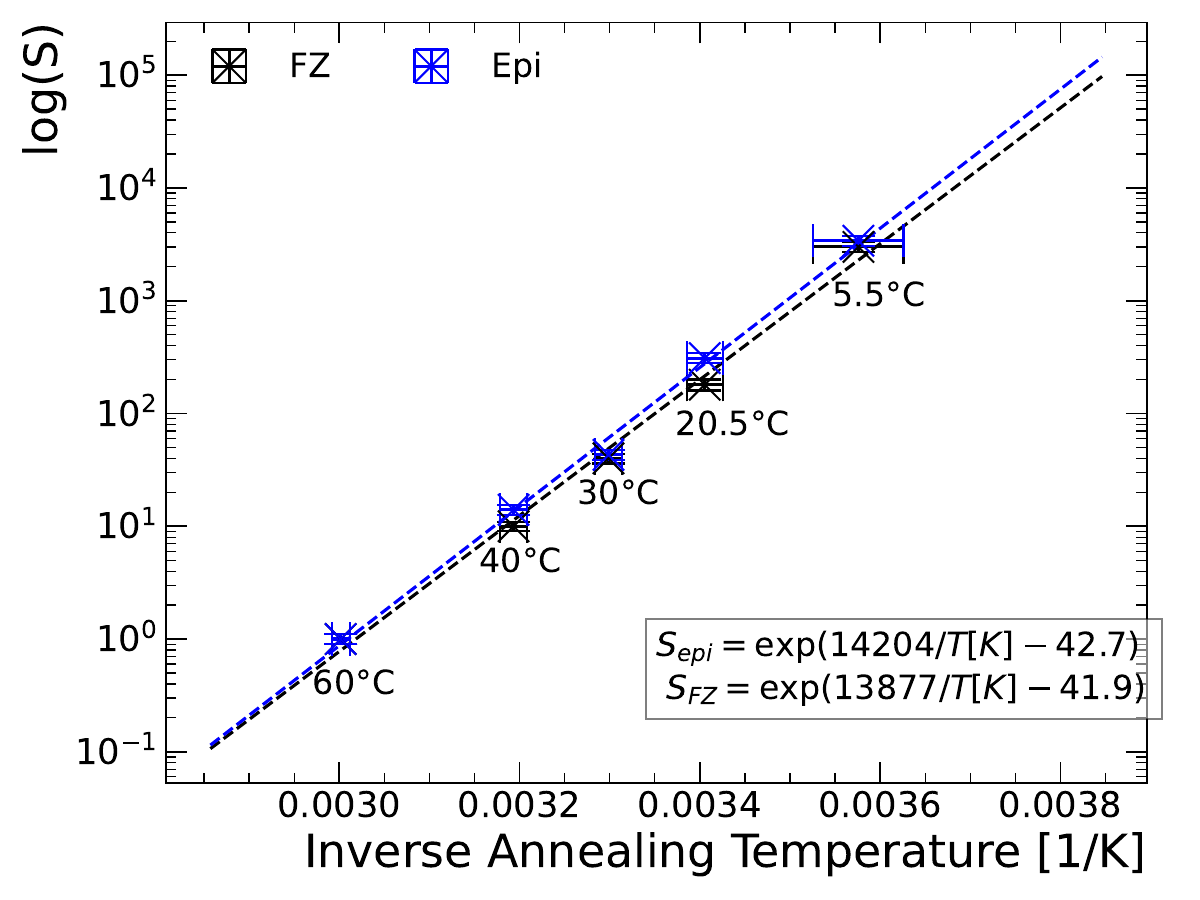}
        \caption{Extracted scaling factors in dependence of the annealing temperature, the lines show the applied fit to extract the scaling equations.  \vspace{0.57cm}}
       
        \label{fig:Scaling_leak}
    \end{subfigure}
\caption{Temperature dependence of the annealing time of the leakage current.}
\end{figure}

Figure~\ref{Fig:Leakage} shows that the lower the annealing temperature, the longer the annealing times. To get a rough idea of the temperature dependence, the annealing times of the lower annealing temperatures are scaled until the course of the leakage current agrees best with the current measured at $60^\circ$C annealing. This was done only as an estimation by eye, the scaling factors (divisors) found to fit best for each temperature are written in the legend of Figure~\ref{fig:leak_timescaled}. By plotting these scaling factors versus temperature and fitting an equation of the form 
\begin{equation}
    S = \exp{(a/T[\rm K]-b)}
    \label{eq:Scaling}
\end{equation}
predictions at other temperatures can be made, as shown in Figure~\ref{fig:Scaling_leak}. 


\section{Charge Collection Efficiency}
\label{sec:CCE}
\begin{figure}
    \centering
    \begin{subfigure}{0.49\textwidth}
        \includegraphics[width=\textwidth]{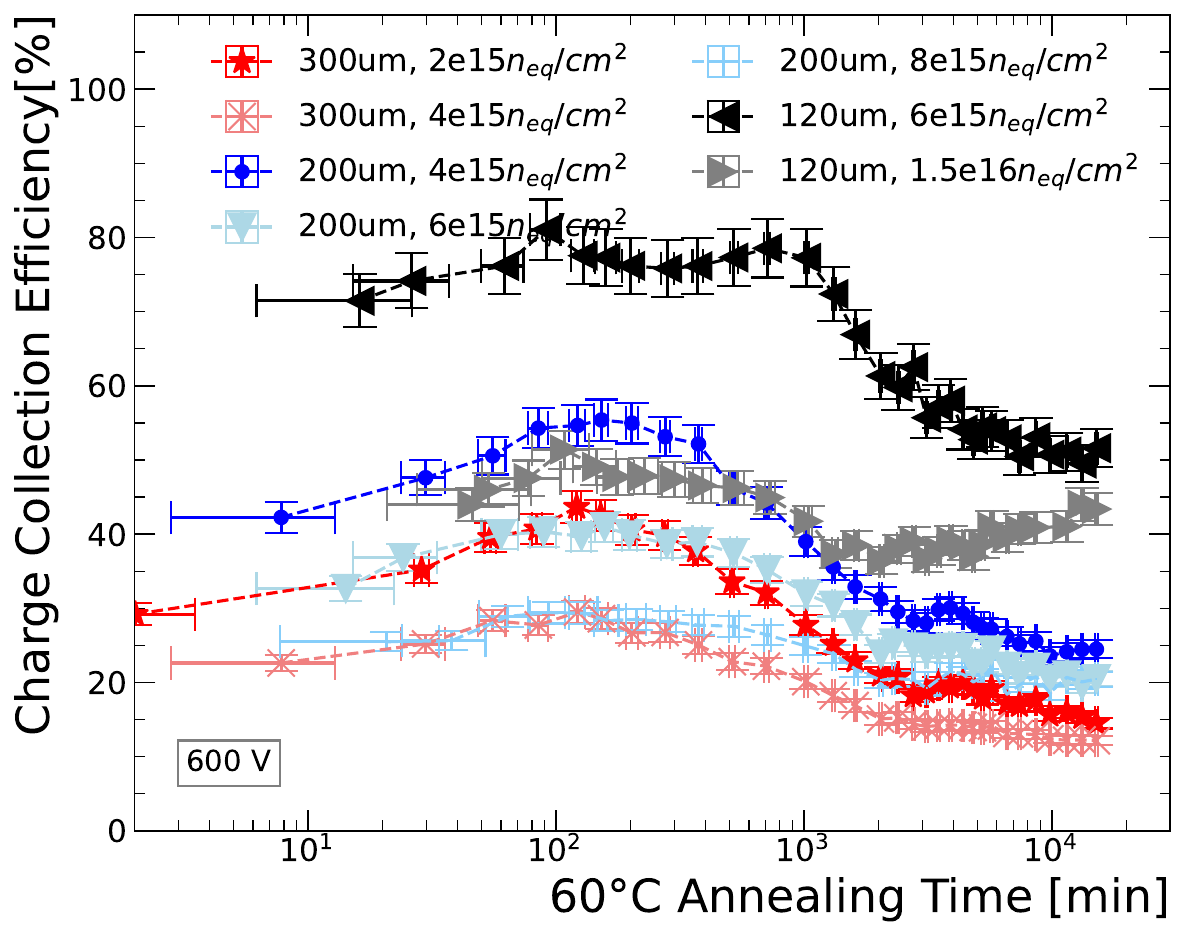}
        \caption{$60^\circ$C annealing}
        \label{fig:CCE_60C}
    \end{subfigure}
    \begin{subfigure}{0.49\textwidth}
        \includegraphics[width=\textwidth]{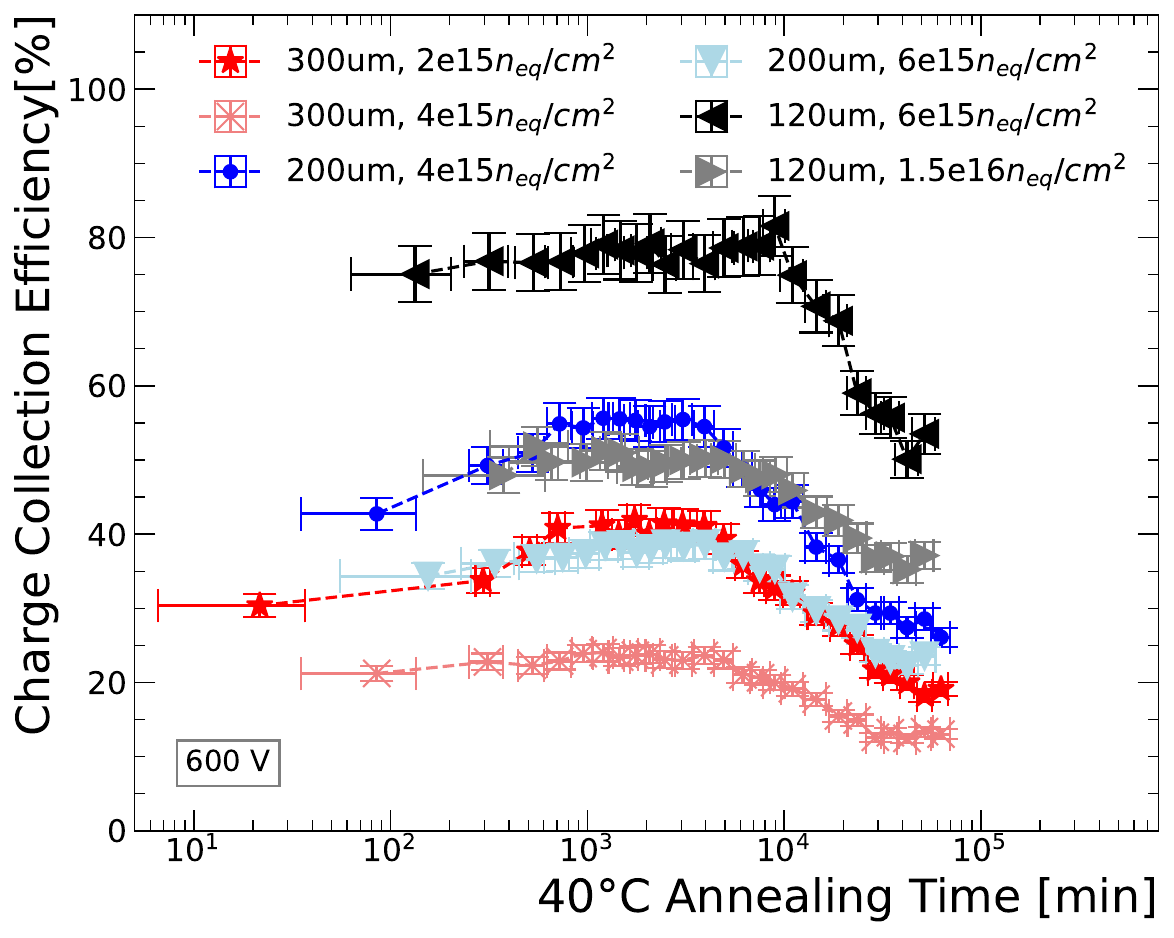}
        \caption{$40^\circ$C annealing}
        \label{fig:CCE_40C}
    \end{subfigure}
        \begin{subfigure}{0.49\textwidth}
        \includegraphics[width=\textwidth]{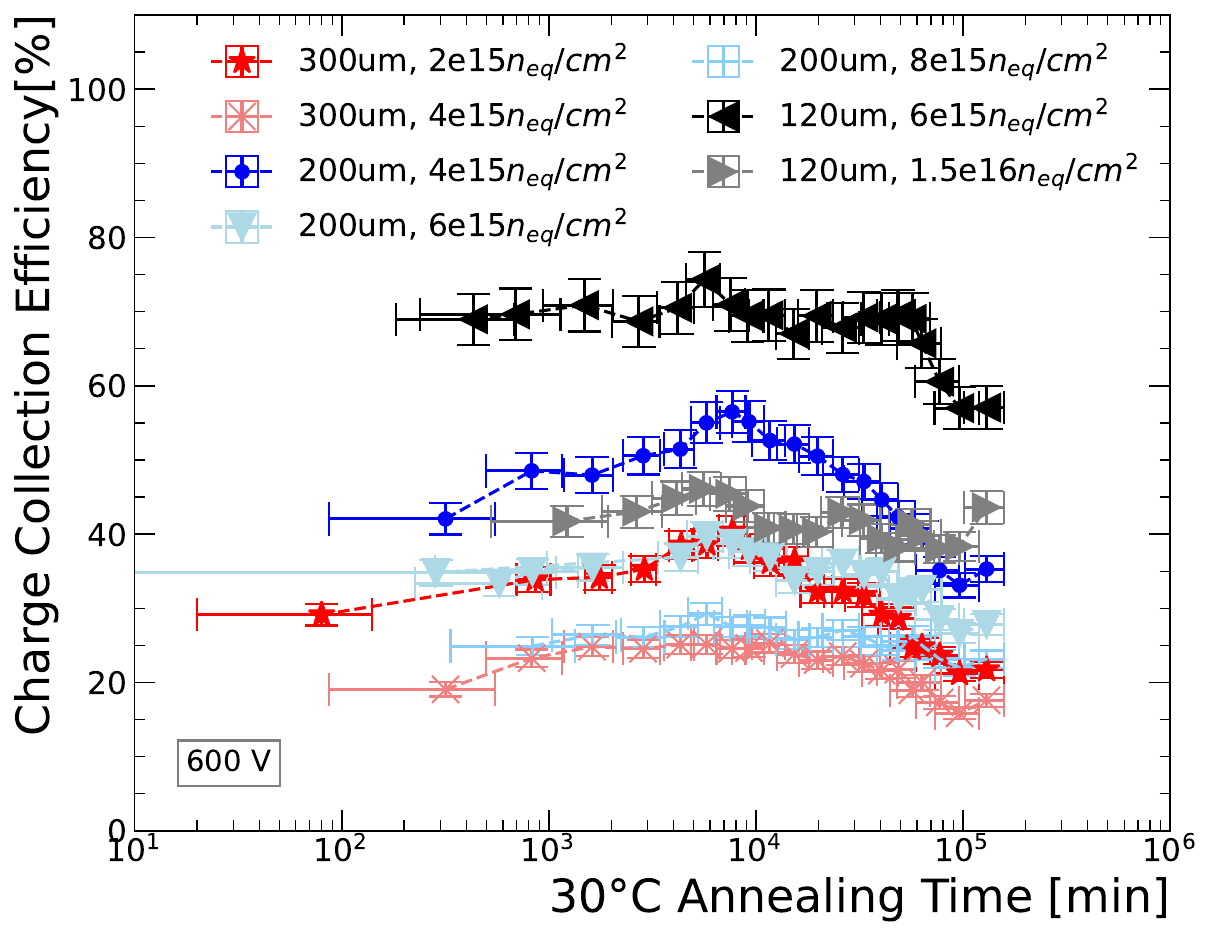}
        \caption{$30^\circ$C annealing}
        \label{fig:CCE_30C}
    \end{subfigure}
    \begin{subfigure}{0.49\textwidth}
        \includegraphics[width=\textwidth]{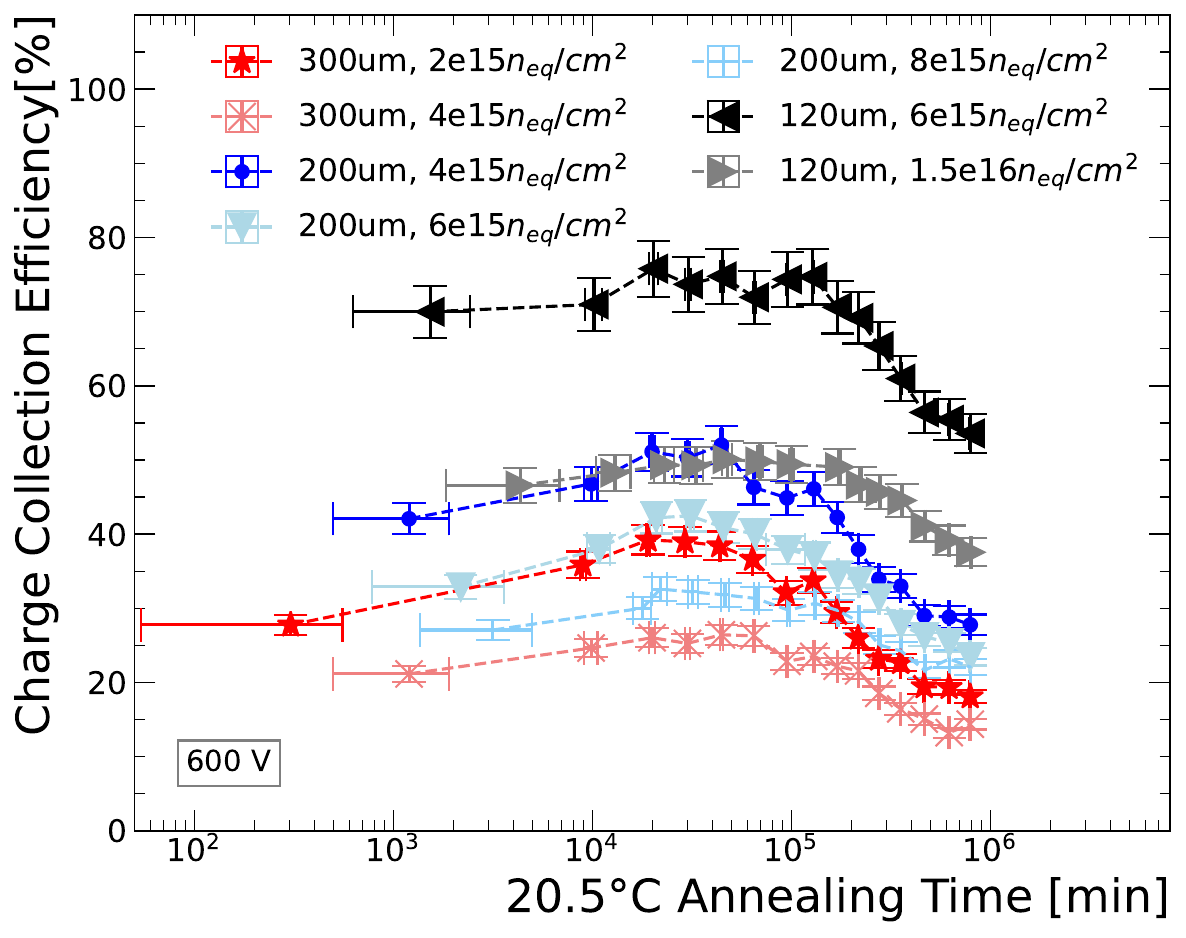}
        \caption{$20.5^\circ$C annealing}
        \label{fig:CCE_RT}
    \end{subfigure}
\caption{Charge collection efficiency at 600\,V in dependence of the annealing time at different temperatures. The dashed lines are shown as eye-guides.}
\label{fig:CCE_Annealing}
\end{figure}

The charge collection of the sensors was measured in 100\,V steps from 0\,V to 900\,V after each annealing step. The charge collection effiency (CCE) was then determined by dividing the measured charge by the charge measured in an unirradiated sensor of the same thickness in the same setup under the same conditions. 
Figure~\ref{fig:CCE_Annealing} presents the CCE measured at the currently foreseen  operation voltage 600\,V of the calorimeter in dependence on the annealing time at $60^\circ$C, $40^\circ$C, $30^\circ$C and $20.5^\circ$C for all different fluences. The dataset at $5.5^\circ$C is only used as reference dataset and will be presented as such later. The general behavior follows the expected course for all sensors: an efficiency increase during the beneficial-annealing dominated times, and a decrease during the long-term annealing, which is dominated by the reverse annealing. 
A closer examination of the time of maximum efficiency reveals a distinct difference between the epitaxial and floatzone sensors. For all temperatures, the maximum is reached earlier for the epitaxial sensors than for the floatzone sensors, e.g., after around 100~minutes at $60^\circ$ C for epitaxial material, while it takes about 150~minutes for the floatzone material.
The relative increase during beneficial annealing is significantly lower for epitaxial sensors as well, especially for the $6\cdot10^{15}\,\rm n_{eq}/cm^{2}$ sensor. For these sensors there is a distinct difference visible to the other sensors as well - an almost constant CCE beyond the maximum, partially resembling almost a second local maximum. Both phenomena can be explained by the fact that these are the only sensors that still exhibit an electric field throughout the entire thickness of the sensor at 600\,V ($V_{\rm sat} <  600V$). Only after annealing times above e.g. 1000\,min at $60^\circ$C this high electric field region decreases, resulting in a significantly stronger decrease of the CCE. In all other sensors the high electric field region is not extending through the entire thickness for all annealing times at 600\,V, leading to this stronger decrease of CCE immediately as the reverse annealing dominates. 

\begin{figure}
    \centering
    \begin{subfigure}{0.49\textwidth}
        \includegraphics[width=\textwidth]{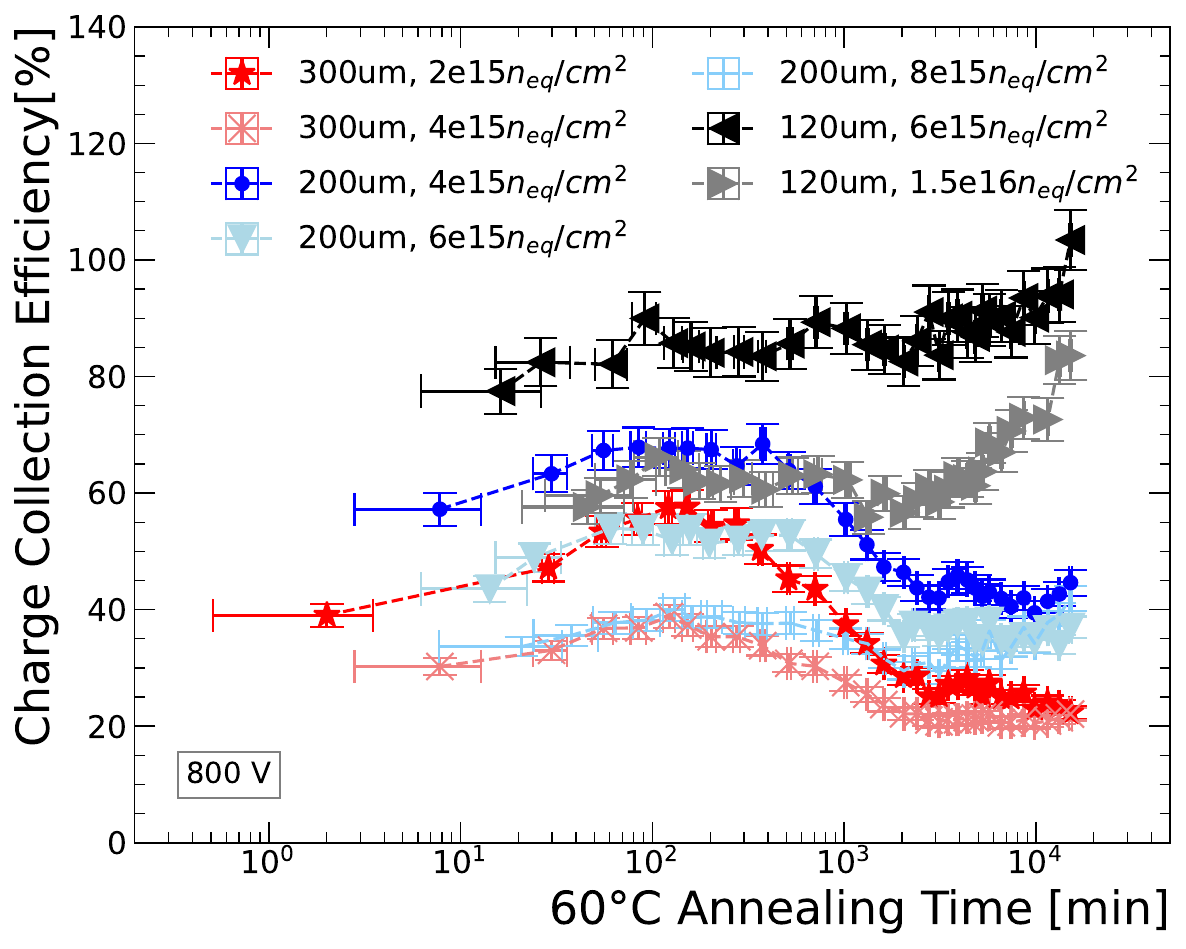}
        \caption{Charge collection efficiency measured at 800\,V in dependence of the $60^\circ$C annealing time.}
        \label{fig:CCE_800V}
    \end{subfigure}
    \begin{subfigure}{0.49\textwidth}
        \includegraphics[width=\textwidth]{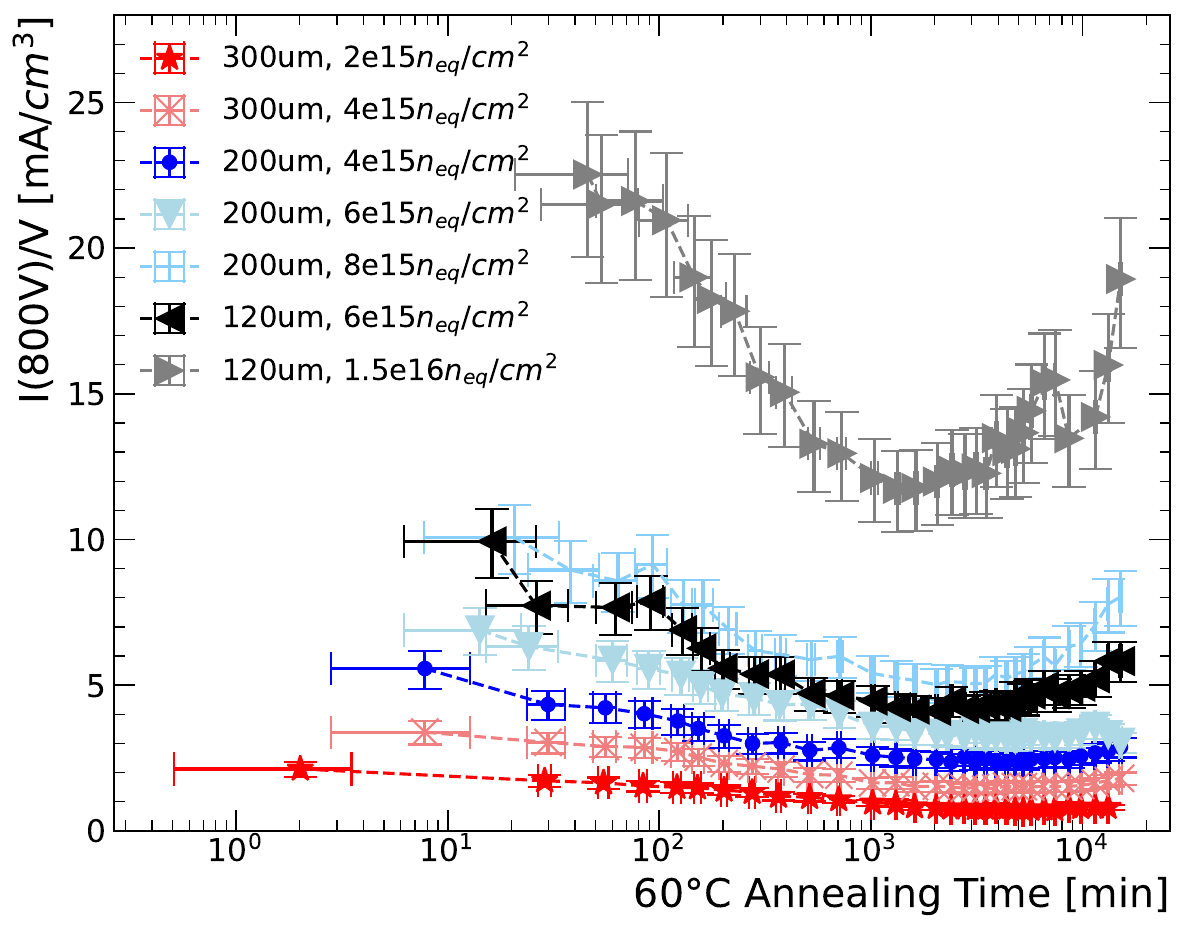}
        \caption{Volume normalised leakage current measured at 800\,V in dependence  of the $60^\circ$C annealing time.}
       
        \label{fig:Ileak_800V}
    \end{subfigure}
\caption{Charge collection efficiency and leakage current at 800\,V as a function of $60^\circ$C annealing time, exhibiting charge multiplication beyond 1000\,min of annealing.  The dashed lines are shown as eye-guides.}
\end{figure}

For the sensors irradiated to the highest fluence of $1.5\cdot10^{16}\,\rm n_{eq}/cm^{2}$ an increase in CCE is visible after annealing times longer than 1000\,min at $60^\circ$C, and slightly for times beyond 5000\,min at $40^\circ$C as well. 
This increase can be explained by the onset of so-called charge multiplication, an effect that has been observed in highly irradiated sensors after long annealing times \cite{mikuvz2011study,diehl2020investigation}. With the increase of the electric field strength close to the junction a critical electric field is reached, leading to impact ionization and thus to an increase in the collected charge. 
For lower annealing temperatures this effect is not visible, as the annealing is significantly slower. 
At higher voltages, the effect is even stronger and can be seen for several sensors, as shown in Figure~\ref{fig:CCE_800V} for 800\,V. From 1000\,min onward the efficiency starts to rise for the first sensor. The lower the fluence and the larger the thickness, the later the onset of multiplication. The charge multiplication also affects the leakage current, which increases along with the collected charge, as shown in Figure~\ref{fig:Ileak_800V}, leading to a higher power consumption and a higher noise level. The difference in the strength of the effect and the onset time is reflected in the current as well.

It has to be mentioned that the timescale after which this effect starts to occur is far beyond the amount of annealing expected in HGCAL until the end of HL-LHC.

\section{Effective Doping Concentration}
\label{sec:Neff}
\begin{figure}
    \centering
    \begin{subfigure}{0.48\textwidth}
            \includegraphics[width=\textwidth]{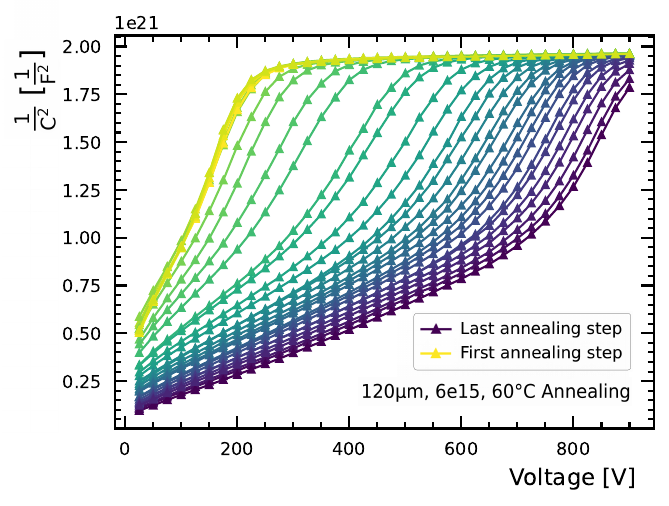}
        \caption{$1/C^2$ as a function of bias voltage for all annealing times.}
        \label{fig:CV-curves}
    \end{subfigure}
    \hfill
    \begin{subfigure}{0.48\textwidth}
  \includegraphics[width=\textwidth]{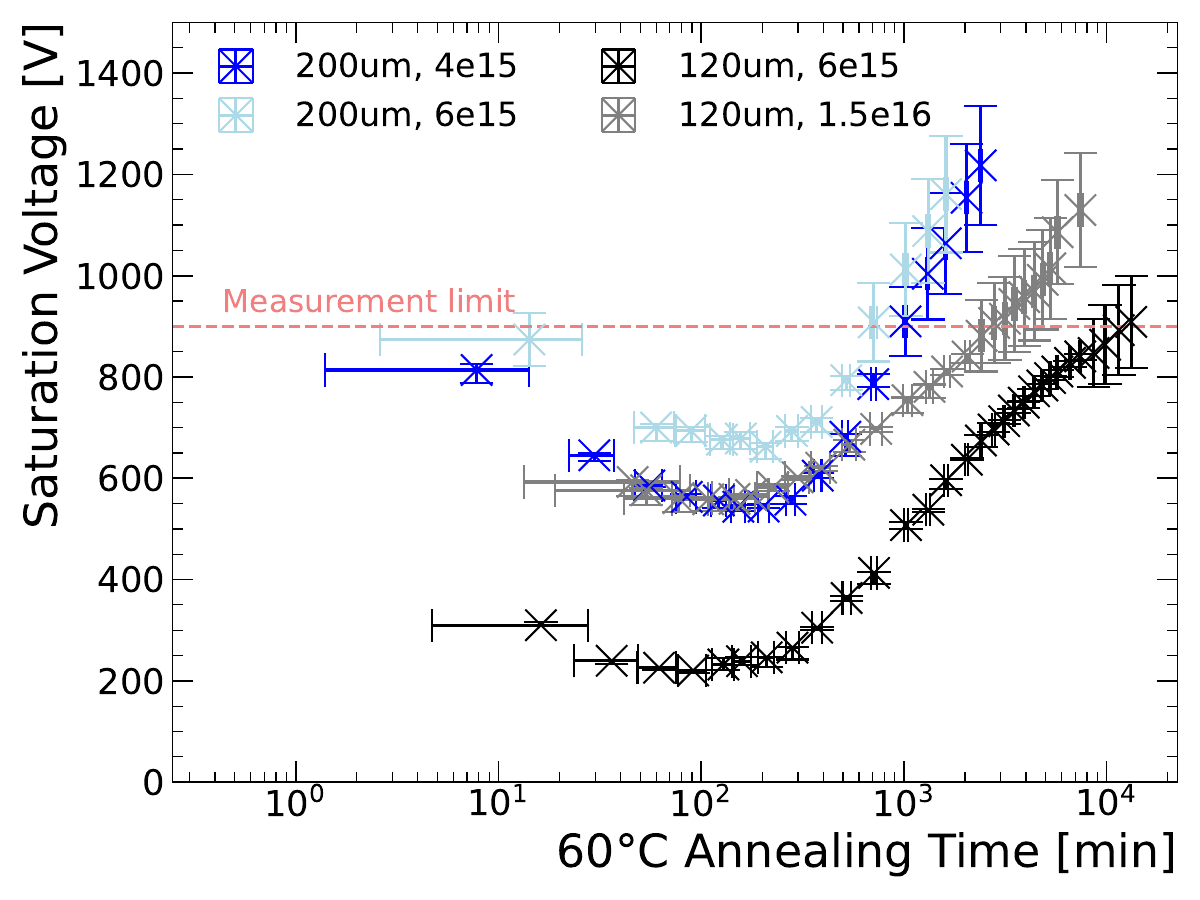}
        \caption{Saturation voltage in dependence of $60^\circ$C annealing time.}
        \label{fig:V_sat_60C}
    \end{subfigure}
\caption{Capacitance measurements of the epitaxial sensor irradiated to $6\cdot10^{15}\rm n_{eq}/cm^{2} $ and saturation voltages as function of annealing time of the sensors annealed at $60^\circ$C.}
\label{Fig:Saturation_voltages}
\end{figure}

The effective doping concentration is calculated from the saturation voltage, which is extracted from the CV measurements. As the CV measurements are temperature and frequency dependent for irradiated sensors, the saturation voltage varies depending on the measurement settings, and thus the absolute value of the extracted effective doping concentration changes. However, since the same frequency and temperature are used for all measurements, the relative development of the extracted effective doping concentration with increasing annealing time can be characterized, since the  values extracted at 2\,kHz and -$20^\circ$C are still a quantity that reflect the value of the effective doping concentration. 
Figure~\ref{fig:V_sat_60C} exemplary shows the course of the saturation voltage in dependence of annealing time for the epitaxial sensor irradiated to $6\cdot10^{15}\rm n_{eq}/cm^{2} $, alongside the recorded $1/C^2$ in dependence of voltage for all annealing times in Figure~\ref{fig:CV-curves}. The maximum bias voltage is 900\,V, therefore, the saturation voltages above 800\,V are not accessible by fitting both the rising and the constant part and calculating the intersection. 
However, the CV curves show that the capacitance beyond saturation remains constant throughout the entire annealing process. This means, assuming the capacitance beyond saturation remains constant also for longer annealing times, that the saturation voltages can be extracted by fitting only the rising slope and calculating the intersection with the expected saturation capacitance known from all previous annealing steps.   In Figure~\ref{fig:CV-curves} it is visible that there are two distinct slopes in the CV-curves, thus, this method is only reliable as long as both slopes are clearly separable, since the slow rising first slope would lead to a great overestimation of the saturation voltage, which limited the extraction to a voltage range between 800\,V  and about 1200\,V. 
The end-capacitance assumption method is considered less precise, therefore a larger uncertainty of $10\%$ is assumed. 

The effective doping concentration is calculated using the equation 
\begin{equation}
    N_{\rm eff} = \frac{2\varepsilon \varepsilon_0}{q_0}\frac{V_{\rm sat}}{d^2}
\end{equation}
where $\varepsilon_0$ is the dielectric constant in vacuum ($8.85\cdot10^{-14}F/$cm), $\varepsilon$ the material-dependent permittivity modifier (11.68 for silicon), $q_0$ the charge of one electron in Coulomb ($1.6 \cdot 10^{-19}$C), and $d$ the active thickness of the diode. 

The Hamburg model provides a mathematical description of its dependence on annealing \cite{Moll:1999kv}. It states that the effective doping concentration due to radiation damage for equivalent annealing times of the order of minutes or longer at $60^\circ$C is made up of three parts. 
First, a material dependent stable damage, represented by a term that can be approximated as 
\begin{equation}
\label{eq:gc}
    N_C = g_c \Phi
\end{equation}
for sensors with a high resistivity in the order of $k\Omega$cm \cite{Moll:1999kv}. 
Second, the beneficial annealing described by the term 
\begin{equation}
\label{eq:g_a}
    N_A = g_a \Phi \exp{(-t/\tau_a)}\, ,
\end{equation}
with $t$ being the annealing time and $\tau_a$ being the beneficial annealing time constant. 
Third, the reverse annealing modeled by the term 
\begin{equation}
\label{eq:g_y}
    N_\gamma = g_\gamma\Phi\left(1-\frac{1}{1+t/\tau_\gamma} \right)
\end{equation}
which is described by an upper limit $g_\gamma$ and a time constant $\tau_\gamma$.

In this study, $g_a, g_\gamma, g_c, \tau_a$ and $\tau_\gamma$ are considered free parameters. Floatzone and epitaxial sensors are treated separately, as a certain material dependence can be expected in the stable damage. The results for the Floatzone sensors in the following include only the 200\,\textmu m thick samples.

\begin{figure}
    \centering
    \begin{subfigure}{0.49\textwidth}
        \includegraphics[width=\textwidth]{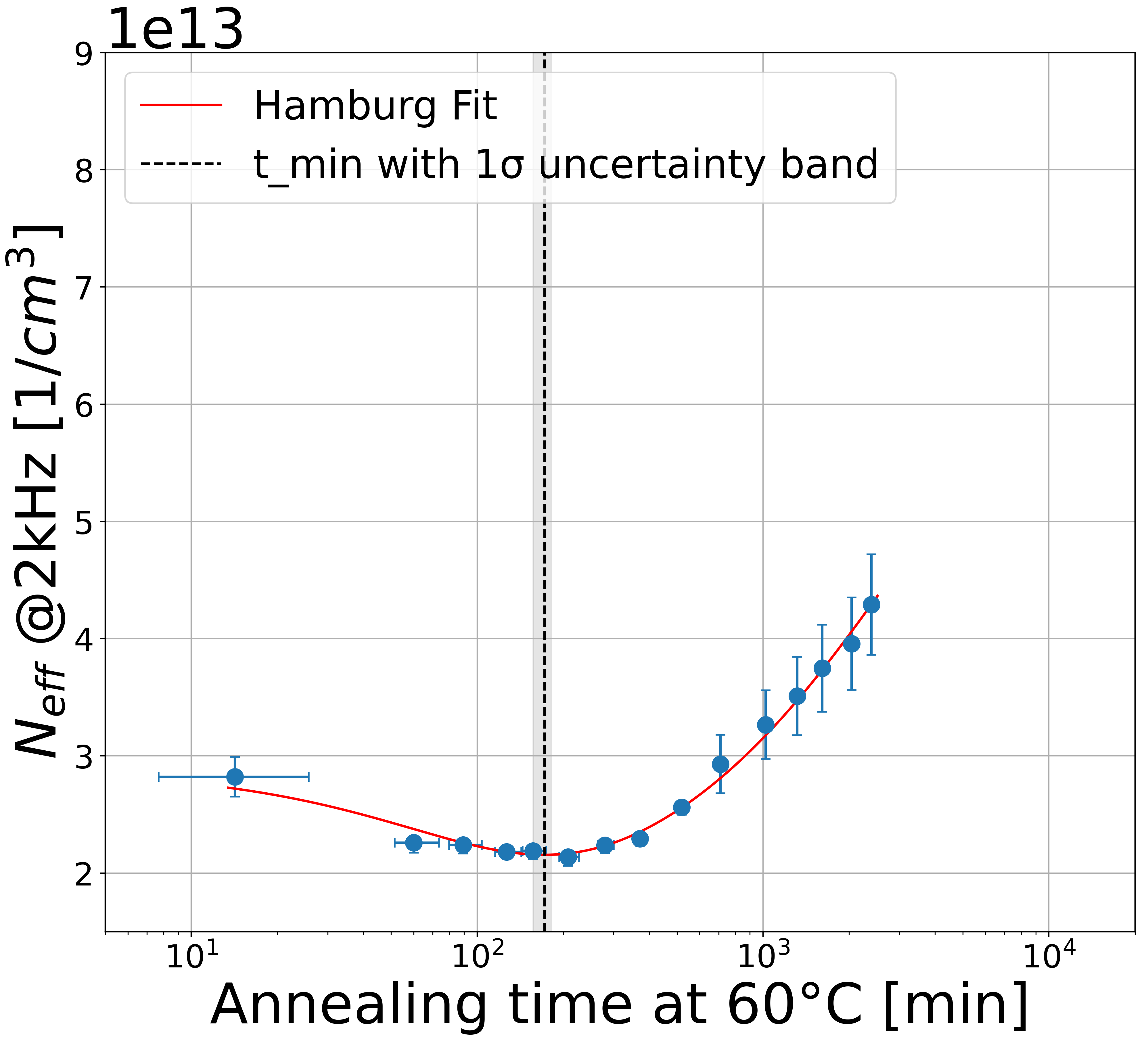}
        \caption{$200\,\upmu$m FZ sensor.}
        \label{fig:Fit_6e15FZ}
    \end{subfigure}
    \begin{subfigure}{0.49\textwidth}
        \includegraphics[width=\textwidth]{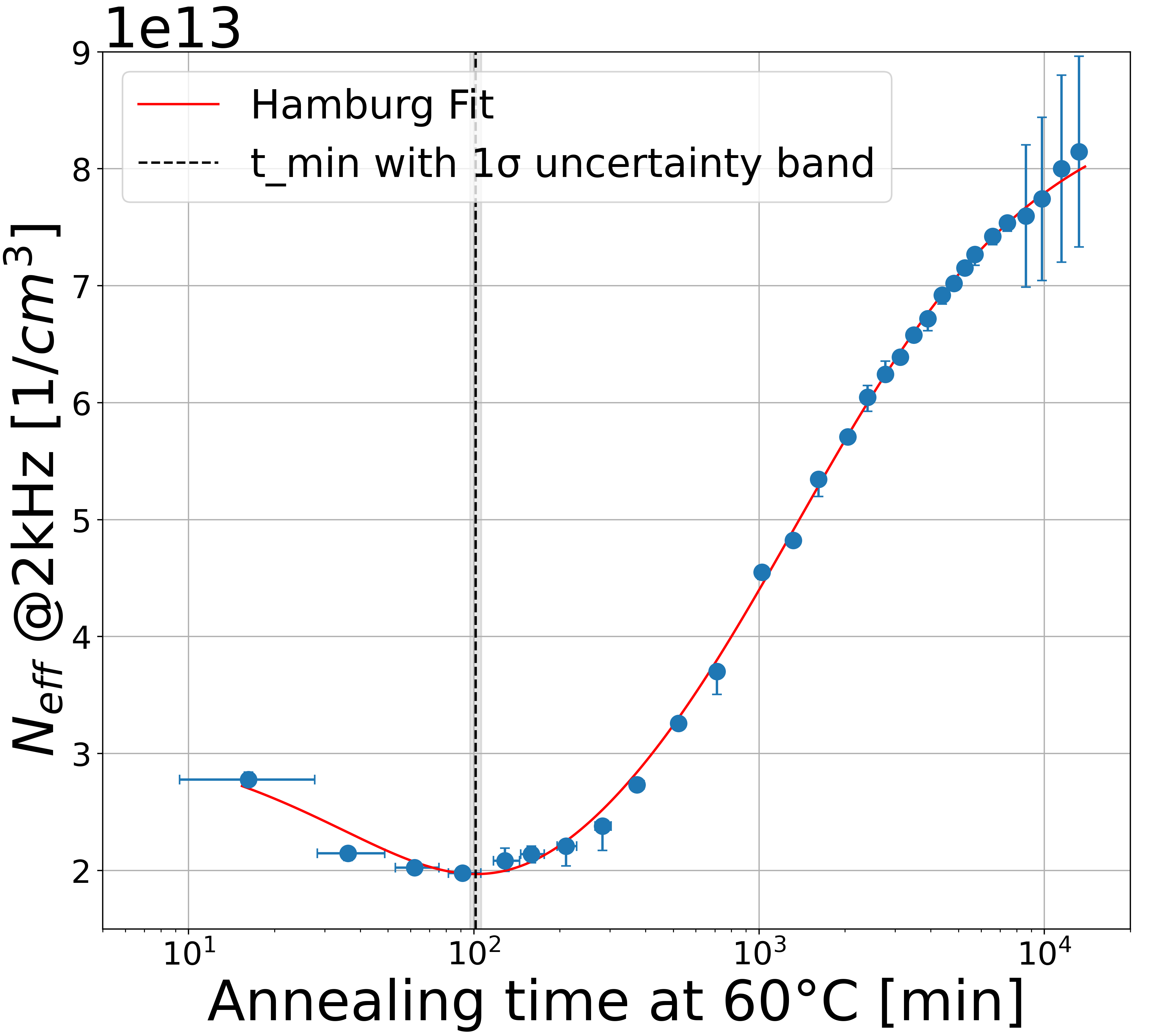}
        \caption{$120\,\upmu$m EPI sensor.}
       
        \label{fig:Fit_6e15_EPI}
    \end{subfigure}
\caption{Hamburg model fit for sensors irradiated to $6\cdot10^{15}\,\rm n_{eq}/cm^{2}$ annealed at $60^\circ$C.}
\label{Fig:Hamburg_Fits}
\end{figure}

As can be seen in Figure~\ref{Fig:Hamburg_Fits} for the sensors irradiated to a fluence of $6\cdot10^{15}\,\rm n_{eq}/cm^{2}$ annealed at $60^\circ$C, the Hamburg model fits provide a good description for both the $200\,\upmu$m thick FZ sample (a), and the $120\,\upmu$m thick EPI sample (b). 
For the thinner sensor significantly more data points are available in the reverse-annealing dominated region, increasing the reliability of the fits significantly. The lack of data for FZ samples after long annealing times is due to the extracted saturation voltages starting to exceed the measurement limits. This is also the reason why the $300\,\upmu$m thick sensors and the $200\,\upmu$m $8\cdot10^{15}\,\rm n_{eq}/cm^{2}$ sensors could not be used for this part of the analysis.

\paragraph{\textbf{ Minimum in $\mathbf{\rm N_{eff}}$}}

The first extracted parameter to evaluate from the fits is the timing of the minimum. From the reference parameters of the Hamburg model, the minimum is expected after about 80 minutes of $60^\circ$C annealing. 
In Table~\ref{tab:Neff} the timing of the minimum is listed for all sensors. 
The given uncertainties origin from the uncertainties of the fits, as other uncertainties such as the time offset due to the in-reactor annealing and the uncertainties on the effective doping concentration are taken into account in these fits. 

The trend of a later minimum for the $200\,\mu$m floatzone sensors compared to the $120\,\mu$m epitaxial sensors is evident for all annealing temperatures, hinting toward a material dependence of the annealing processes. 
For the floatzone sensors, additionally, a time increase up to almost 30\% is observed for the higher fluence. However, the floatzone sensors  irradiated to $6\cdot10^{15}\,\rm n_{eq}/cm^{2}$ have the least data points available, thus the fits have the highest uncertainties. To confirm a fluence dependence of the timing of the minima more fluence points would be necessary. 
\begin{table}
    \centering
        \caption{Timing of the minima in $N_{\mathrm{eff}}$ in minutes for all annealing temperatures.}
    \begin{tabular}{|c|c|c|c|c|}
    \hline
    Fluence $\rm [n_{eq}/cm^{2}]$& $60^\circ$C [$10^{2}$min] & $40^\circ$C [$10^{3}$min]&$30^\circ$C [$10^{3}$min]& $20.5^\circ$C [$10^{4}$min]\\
         \hline
         $4.0\cdot10^{15}$ FZ & $1.65\substack{+ 0.06\\ -0.05}$& $1.73\substack{+ 0.10\\ -0.11}$&$6.85\substack{+ 0.43\\ -0.47}$& $3.29\substack{+ 0.14\\ -0.18}$\\\hline
         $6.0\cdot10^{15}$ FZ &$1.72\substack{+ 0.09\\ -0.14}$ &$1.97\substack{+ 0.20\\ -0.27}$&$8.95\substack{+ 0.66\\ -0.97}$& $3.94\substack{+0.16\\ -0.17}$ \\\hline
         $6.0\cdot10^{15}$ EPI &$1.02\substack{+ 0.04\\ -0.05}$ &$0.96\substack{+0.06\\ -0.07}$&$3.88\substack{+ 0.35\\ -0.46}$& $1.76\substack{+0.14\\ -0.18}$\\\hline
         $1.5\cdot10^{16}$ EPI &$1.10\substack{+ 0.13\\ -0.15}$ &$1.13\substack{+ 0.15\\ -0.20}$&$3.07\substack{+ 0.52\\ -0.65}$& $1.35\substack{+0.38\\ -0.50}$\\\hline
             \end{tabular}

    \label{tab:Neff}
\end{table}

\begin{figure}
    \centering
        \includegraphics[width=0.75\textwidth]{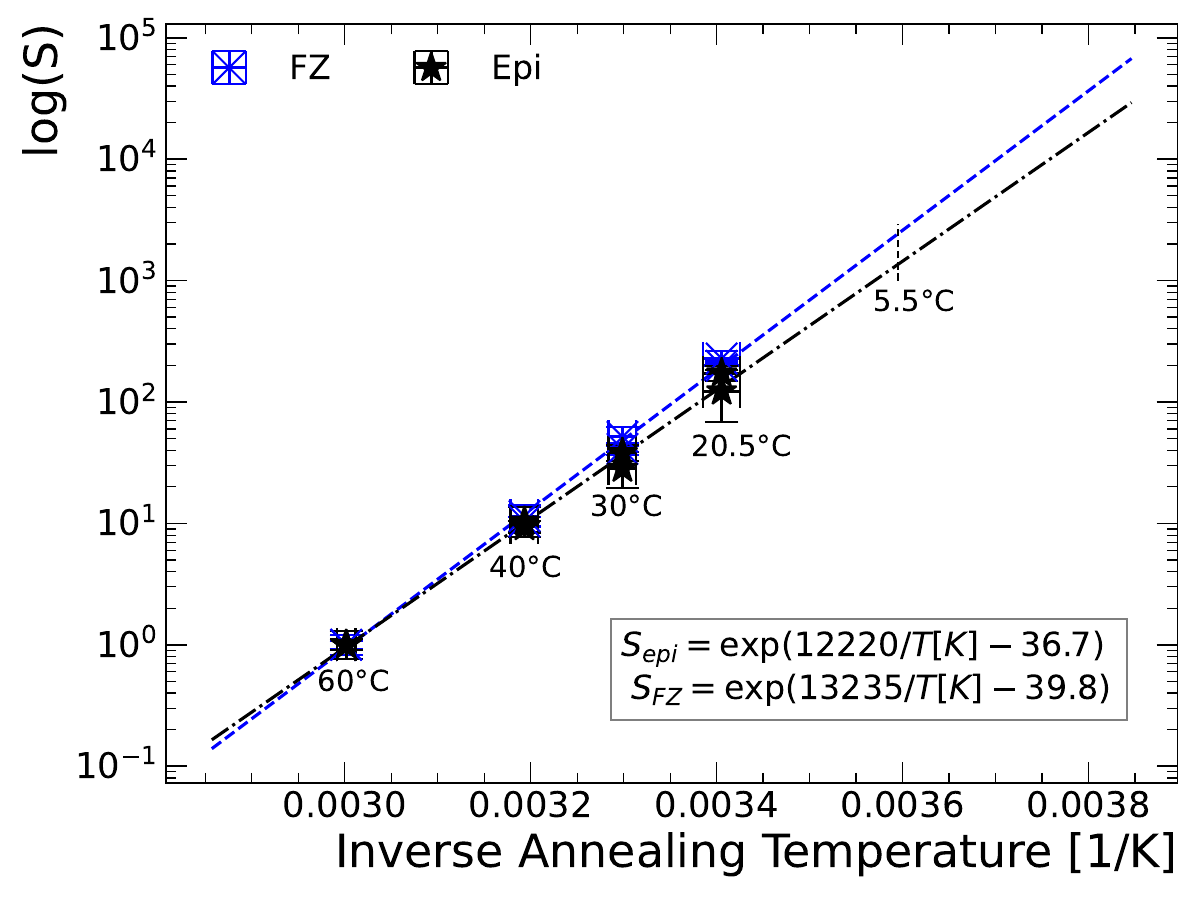}
        \caption{Scaling of the minima in the effective doping concentration in dependence of annealing temperature with respect to $60^\circ$C, the dashed lines represent the fit to extract the scaling equation, the $5.5^\circ$C mark was added to visualize better where the reference measurement set is placed on extrapolated scaling.}
        \label{fig:Scaling_Neff}
    \end{figure}

\begin{figure}
    \begin{subfigure}{0.48\textwidth}
        \includegraphics[width=\textwidth]{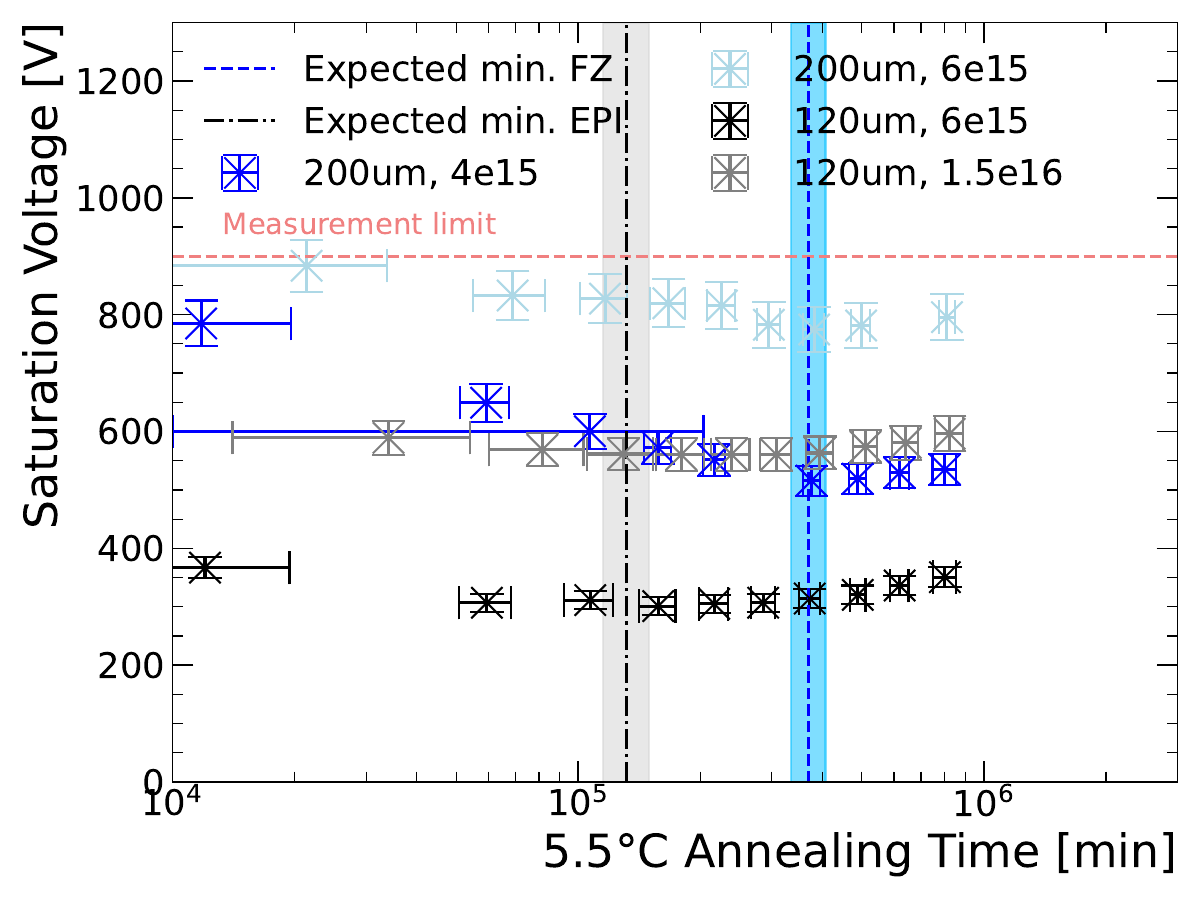}
        \caption{Extracted saturation voltage in dependence of the annealing time.}
       
        \label{fig:Vsat_lowT}
    \end{subfigure}
    \hfill
    \begin{subfigure}{0.48\textwidth}
        \includegraphics[width=\textwidth]{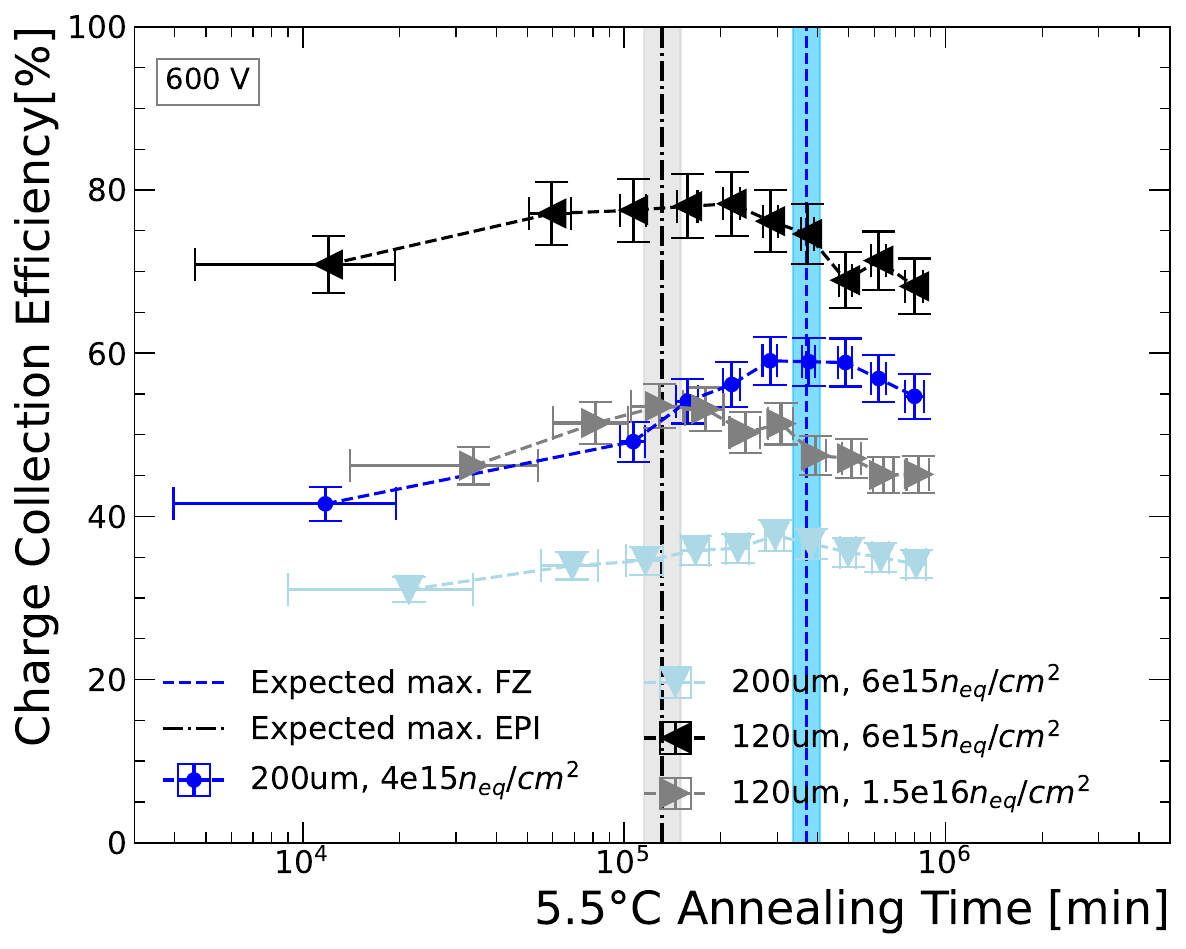}
        \caption{Charge collection efficiency in dependence of the annealing time. The lines act as eye-guides.}
       
        \label{fig:CCE_lowT}
    \end{subfigure}
\caption{Crosscheck of the extracted timing of the minima (maxima for CCE) with the dataset recorded at $5.5^\circ$C annealing. The black vertical line with the grey uncertainty band represents the expected time for EPI sensors, while the blue line with the light blue uncertainty band represents the expected time for FZ sensors.}
\label{Fig:Crosscheck}
\end{figure}

Using the extracted minima, scaling between temperatures becomes possible. By plotting the scaling factors of the minima between the annealing temperatures with respect to $60^\circ$C in dependence of annealing temperature as presented in Figure~\ref{fig:Scaling_Neff}, equation \ref{eq:Scaling} can be fitted and used to calculate the expected timing of the minimum at any given temperature. The fits were done for floatzone and epitaxial sensors separately, as there is a significant difference visible between the two materials. 
The scaling factors were extrapolated to $5.5^\circ$C annealing and the expected minima were calculated as a cross-check with the available dataset, as shown in Figure~\ref{Fig:Crosscheck}. 
The observed minima of the saturation voltage are well in agreement with the extrapolated values, with only small deviations for the individual fluences. The exact minima of the epitaxial sensor data are difficult to determine from the available dataset, as there is not a strong decrease visible through the beneficial annealing and due to the slow annealing at $5.5^\circ$C, a larger number of points lies within the range of the minimum. However, in the charge collection efficiency (Figure~\ref{fig:CCE_lowT}) the increase is slightly better visible, and seems to coincide quite well with the extrapolated value. For the larger fluences, the uncertainty regarding the calculated in-reactor annealing is higher, so a potential time offset due to the calculation of the in-reactor annealing time may have a larger impact than for the lower fluences. 

Since for many detector experiments the annealing times are predominantly determined by the shutdown periods that are expected to be within a time region still affected by both beneficial and reverse annealing, this scaling can be a good indicator of the expected status of a sensor at an annealing temperature that was not previously evaluated. For long annealing times, when the reverse annealing becomes dominant, a different time and temperature dependence has to be considered.

\paragraph{\textbf{Beneficial and reverse annealing time constants}}

The next interesting parameters to evaluate are the beneficial and reverse annealing time constants. They describe how fast the annealing progresses at a given temperature and how their temperature dependence reflects the activation energy. 
The extracted beneficial time constants and the corresponding acceleration factors with respect to $60^\circ$C are summarized in Table~\ref{tab:taos_beneficial}, while the reverse annealing time constants and their acceleration factors are summarized in Table~\ref{tab:taos_reverse}. The acceleration factors are calculated by dividing the annealing time constant at $60^\circ$C by the annealing time constants of the other annealing temperatures, e.g. $A_{a,40} = \tau_{a,60}/\tau_{a,40}$.

\begin{table}
    \centering
        \caption{Beneficial annealing time constants and the corresponding acceleration factors with respect to $60^\circ$C.}
    \begin{tabular}{|p{18mm}|p{15mm}|p{12mm}|p{23mm}|p{23mm}|p{23mm}|}
    \hline
    &$\rm n_{eq}/cm^{2}$& $60^\circ$C & $40^\circ$C &$30^\circ$C & $20^\circ$C \\
         \hline
      $\tau_a$  FZ [min] & $4.0\cdot10^{15}$  \newline $6.0\cdot10^{15}$ & $70\pm10$ \newline $80\pm30$& $(6.5\pm0.9)\cdot10^2$\newline $(7.3\pm4.5)\cdot10^2$&$(2.5\pm0.4)\cdot10^3$\newline $(3.8\pm0.9)\cdot10^3$& $(1.2\pm0.1)\cdot10^4$ \newline$(1.7\pm0.7)\cdot10^4$\\\hline

            $\tau_a$  EPI [min] & $6.0\cdot10^{15}$  \newline $1.5\cdot10^{16}$ & $50\pm10$ \newline $40\pm20$& $(3.7\pm0.7)\cdot10^2$\newline $(3.7\pm1.2)\cdot10^2$&$(1.4\pm0.4)\cdot10^3$\newline $(1.1\pm0.3)\cdot10^3$& $(8.0\pm1.8)\cdot10^4$ \newline$(3.3\pm1.5)\cdot10^4$\\\hline
      Acc. factor FZ $A_a$ & $4.0\cdot10^{15}$  \newline $6.0\cdot10^{15}$ & 1  \newline 1 & $1.1\cdot10^{-1}$\newline $1.1\cdot10^{-1}$ & $2.8\cdot10^{-2}$\newline $2.1\cdot10^{-2}$& $6.0\cdot10^{-3}$\newline $5.0\cdot10^{-3}$\\\hline
      Acc. factor EPI $A_a$ & $6.0\cdot10^{15}$  \newline $1.5\cdot10^{16}$ & 1  \newline 1 & $1.3\cdot10^{-1}$\newline $1.0\cdot10^{-1}$& $3.3\cdot10^{-2}$\newline $3.3\cdot10^{-2}$ &$6.0\cdot10^{-3}$ \newline$1.1\cdot10^{-2}$ \\\hline

             \end{tabular}

    \label{tab:taos_beneficial}
\end{table}

\begin{table}
    \centering
        \caption{Reverse annealing time constants and the corresponding acceleration factors with respect to $60^\circ$C.}
    \begin{tabular}{|p{18mm}|p{15mm}|p{12mm}|p{23mm}|p{23mm}|p{23mm}|}
        \hline
    &$\rm n_{eq}/cm^{2}$& $60^\circ$C & $40^\circ$C &$30^\circ$C & $20^\circ$C \\
         \hline
      $\tau_\gamma$  FZ [h] & $4.0\cdot10^{15}$  \newline $6.0\cdot10^{15}$ & $50\pm20$ \newline $50\pm20$& $(6.4\pm 3.0)\cdot10^2$\newline $(1.7\pm1.0)\cdot10^3$&$(3.1\pm2.0)\cdot10^3$\newline $(7.7\pm6.5)\cdot10^3$& $(1.3\pm0.8)\cdot10^4$ \newline$(1.7\pm0.3)\cdot10^4$\\\hline

            $\tau_\gamma$  EPI [h] & $6.0\cdot10^{15}$  \newline $1.5\cdot10^{16}$ & $20\pm10$ \newline $30\pm10$& $(3.2\pm0.4)\cdot10^2$\newline $(3.6\pm1.1)\cdot10^2$&$(1.5\pm0.3)\cdot10^3$\newline $(2.1\pm1.1)\cdot10^3$& $(5.3\pm0.5)\cdot10^3$ \newline$(9.9\pm1.8)\cdot10^3$\\\hline
      Acc. factor \newline FZ $A_\gamma$& $4.0\cdot10^{15}$  \newline $6.0\cdot10^{15}$ & 1  \newline 1 & $7.4\cdot10^{-2}$\newline $3.2\cdot10^{-2}$ & $1.5\cdot10^{-2}$\newline$7.0\cdot10^{-3}$ & $3.6\cdot10^{-3}$ \newline $3.2\cdot10^{-3}$ \\\hline
      Acc. factor \newline EPI $A_\gamma$& $6.0\cdot10^{15}$  \newline $1.5\cdot10^{16}$ & 1  \newline 1 & $7.1\cdot10^{-2}$ \newline $7.5\cdot10^{-2}$ &$1.5\cdot10^{-2}$\newline $1.3\cdot10^{-2}$& $4.2\cdot10^{-3}$ \newline$2.7\cdot10^{-3}$ \\\hline

             \end{tabular}

    \label{tab:taos_reverse}
\end{table}

\begin{figure}
    \centering
    \begin{subfigure}{0.48\textwidth}
        \includegraphics[width=\textwidth]{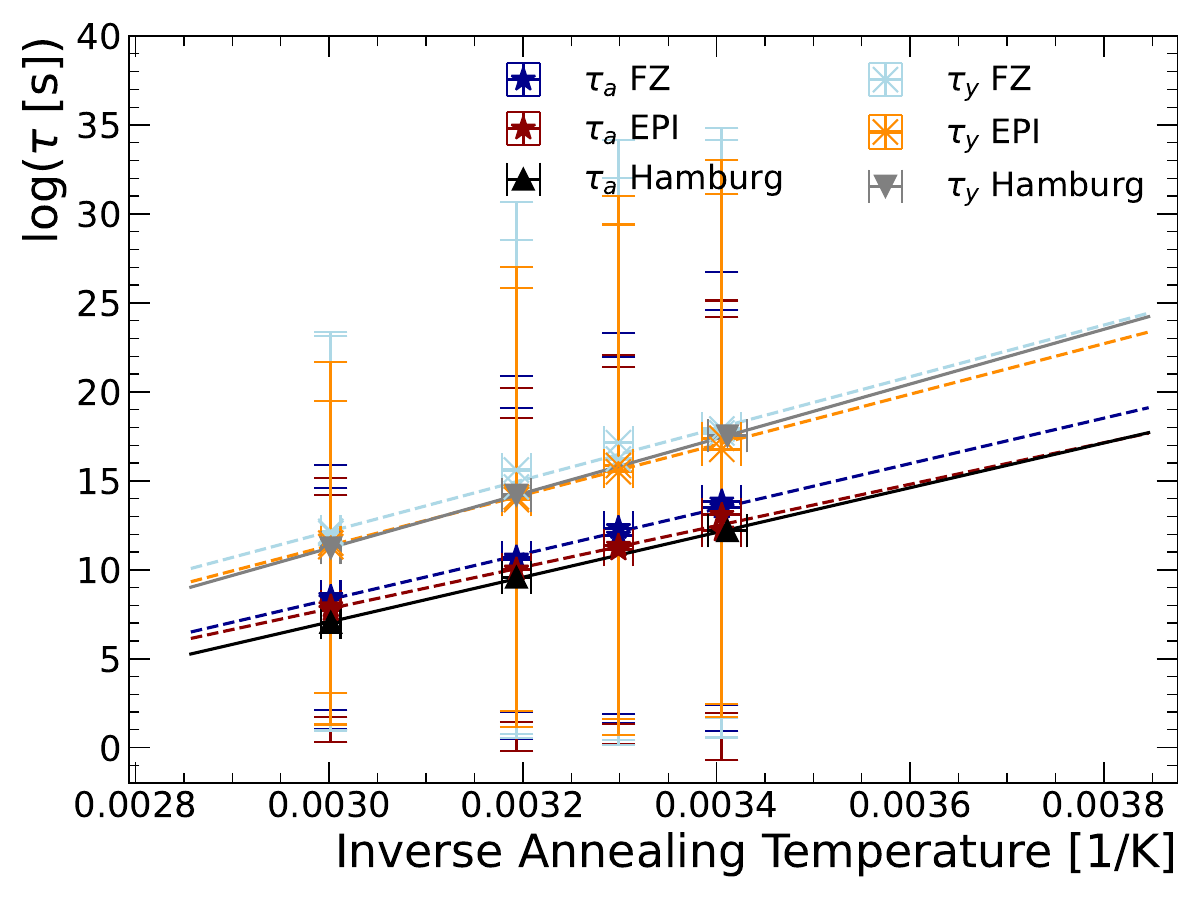}
        \caption{Beneficial ($\tau_a$) and reverse ($\tau_\gamma$) annealing time constants in dependence of the annealing temperature.}
        \label{fig:Time_Constants}
    \end{subfigure}
    \hfill
    \begin{subfigure}{0.48\textwidth}
        \includegraphics[width=\textwidth]{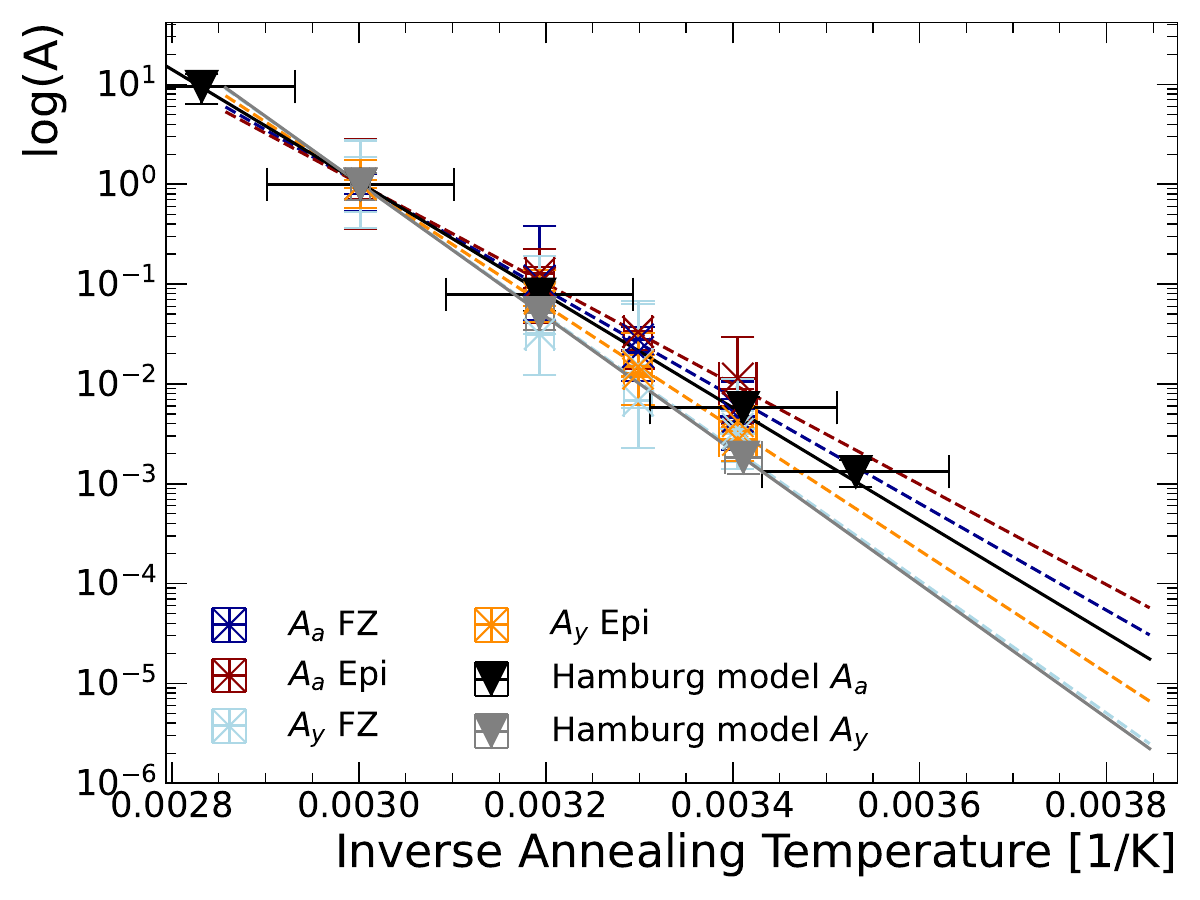}
        \caption{Acceleration factors with respect to $60^\circ$C annealing as a function of the annealing temperature.}
       
        \label{fig:Acceleration}
        \vspace{0.3cm}
    \end{subfigure}
\caption{Annealing time constants and their acceleration factors in dependence of the annealing temperature. The dashed lines represent the temperature dependent fits.}
\label{Fig:taos}
\end{figure}

In Figure~\ref{Fig:taos} the annealing time constants (a) and the acceleration factors (b) are shown in dependence of annealing temperature, alongside the reference values of the Hamburg model. By fitting $\log(\tau)$ and $\log(A)$  with a linear function, the activation energies as well as the scaling factors from a given temperature to $60^\circ$C can be extracted. 
The fits were done for the floatzone sensors and the epitaxial sensors separately, as significant differences between the materials were observed.

From the beneficial and reverse annealing time constants, the two activation energies $E_a$ can be extracted from 
\begin{equation}
    \frac{1}{\tau} = k_0\exp{\left(-\frac{E_a}{k_BT_a}\right) }~,
\end{equation}
where $k_0$ is the frequency factor, $T_a$ is the annealing temperature, and $k_B$ the Boltzmann constant. 
The extracted activation energies are summarized in Table~\ref{tab:Activation_energy} for both beneficial and reverse annealing, along with the reference values of the Hamburg model. The uncertainty of the reverse annealing activation energy for the floatzone sensors is significantly larger, which is caused by the large uncertainties of the fits to the sensors irradiated to $6\cdot10^{15}\,\rm n_{eq}/cm^{2}$ due to the lack of data points in the reverse annealing dominated region caused by the saturation voltages significantly exceeding the measurement limits early on.

\begin{table}[]
    \caption{Activation energies of beneficial and reverse annealing for the two different sensor materials. The Hamburg model values are added for comparison.}
    \centering
    \begin{tabular}{|c|c|c|}
    \hline
        & $E_{a,a}$ [eV] &$ E_{a,\gamma}$ [eV]  \\\hline
     Floatzone   & $1.10 \pm 0.04$ & $1.25 \pm 0.08$  \\\hline
     Epitaxial  & $1.01 \pm 0.04$ & $1.22 \pm 0.05$  \\\hline
     Hamburg model &$1.09\pm0.03$ & $1.33\pm0.03$ \\\hline
    \end{tabular}

    \label{tab:Activation_energy}
\end{table}

Taking a closer look at the annealing time constants presented in Figure~\ref{fig:Time_Constants}, there is an offset for both the beneficial and reverse annealing time constants measured for the floatzone sensors in this campaign compared to the Hamburg model, as they are higher for the entire temperature range presented.  Higher time constants mean that at any given temperature within the range considered, the annealing progresses slower than previously assumed by the Hamburg model. This is in agreement with the observed later minima of the effective doping concentration for the floatzone sensors compared to the Hamburg model. However, the difference in the activation energy for the reverse annealing suggests that compared to the $60^\circ$C annealing, the processes are faster at lower temperatures than assumed by the Hamburg model. 
This affects the calculations of the equivalent annealing times at different temperatures, e.g. the in-reactor annealing times, as well as the expected amount of annealing during shutdown periods. 

For the epitaxial sensors, the offset is present as well, however, it vanishes for lower temperatures due to the significant difference in the temperature dependence of the reverse annealing. While the beneficial annealing is slower throughout the entire temperature range, the reverse annealing at temperatures below roughly $50^\circ$C is processing faster than expected from the Hamburg model. 

A fit equation to extract the acceleration factor $A (T)$ between a given annealing temperature $T$ to $60^\circ$C of the form 
\begin{equation}
    A (T)= \exp{\left(-a/T[K]+b\right) }
    \label{eq:acceleration}
\end{equation}
can be applied to the acceleration factors for FZ and EPI in dependence of the annealing temperature (Figure~\ref{fig:Acceleration}). 
The four extracted parameters for beneficial and reverse annealing are summarized in Table~\ref{tab:Fit_params}. Using the parameters with equation \ref{eq:acceleration}, the acceleration factor for any annealing temperature with respect to $60^\circ$C can be calculated. This can be used to calculate the equivalent time at $60^\circ$C $t_{60}$ for any given time span at another temperature $t_T$ by multiplying the annealing time with the retrieved acceleration factor: $t_{60}=t_T\cdot A(T)$. To calculate the time needed to reach the same annealing effects as during a certain time span at $60^\circ$C, the time span has to be divided by the acceleration factor: $t_T=t_{60}/A(T)$. 

\begin{table}[]
    \centering
        \caption{Extracted fit parameters from Equation \ref{eq:acceleration} for beneficial and reverse annealing acceleration factors.}
    \begin{tabular}{|c|c|c|c|c|}
    \hline
      & Beneficial FZ & Beneficial EPI & Reverse FZ & Reverse EPI\\\hline
       $a$ &  $12314 \pm 299$ & $11569 \pm 217$ &$ 15280 \pm 680$ &$14120\pm213$\\\hline
        $b$ & $37.0\pm0.9$ &$34.7\pm0.7$ &$45.9\pm2.0$ &$42.4\pm0.6$\\\hline
    \end{tabular}

    \label{tab:Fit_params}
\end{table}

To re-calculate the in-reactor annealing times, the equations for the beneficial annealing have been used after the first round of analysis to correct the in-reactor annealing times, which is feasible since the irradiation times are limited and the beneficial annealing dominates in this short term time region.
First, the equivalent time at $60^\circ$C  was calculated for each temperature step during irradiation using an exemplary temperature recording and summed up to the total in-reactor annealing time.  These values  were then used to directly calculate the equivalent times for the other annealing temperatures.
 Besides using the scaling factors, the equivalent annealing time $t_{eq}$ can be calculated using the equation 
\begin{equation}
    t_{eq} = \int_{0}^{t_{end}}\exp{\left[-\frac{E_a}{k_B}\left(\frac{1}{T(t)}-\frac{1}{T_{ref}}\right)\right]}dt
\end{equation}
using the extracted beneficial annealing activation energies $E_a$(see Table \ref{tab:Activation_energy}). $T$ is the temperature the annealing is occurring, $T_{ref}$ the reference temperature the equivalent time is calculated for and $k_B$ is the Boltzman constant. Both methods result in the same values within the uncertainties.

These corrections of the in-reactor annealing times have an effect especially on the timing of the minima. The values calculated using the scaling factors of the Hamburg model and the values calculated using the newly extracted parameters are shown in Table~\ref{tab:Re-calculation} exemplary for the lowest annealing temperature of $5.5^\circ$C, as there the largest differences are present due to the difference in temperature dependence especially for the epitaxial sensors. Due to the difference in temperature dependence, the equivalent times at the annealing temperatures below $30^\circ$C are shorter than assumed, while the times at $40^\circ$C remain very similar and the times at $60^\circ$C are slightly longer. 

\begin{table}
    \centering
        \caption{Equivalent in-reactor annealing times exemplary for $5.5^\circ$C calculated using the Hamburg model values (no distinction between FZ and epi) and the corrected values using the scaling parameters of this campaign for FZ and epi sensors separately to illustrate the effect of the corrections.}
    \begin{tabular}{|P{30mm}|P{40mm}|P{45mm}|}
    \hline
    $\rm n_{eq}/cm^{2}$ & Est. initial $5.5^\circ$C [min] & Corr. est. initial $5.5^\circ$C [min]\\
         \hline
          $2.0\cdot10^{15}$ FZ & $(3.19\substack{+2.81 \\ -1.43})\cdot10^3$& $(2.99\substack{+2.59 \\ -1.42})\cdot10^3$\\\hline
         $4.0\cdot10^{15}$ FZ&$(1.27\substack{+1.06 \\ -0.50})\cdot10^4$ &$(1.18\substack{+0.97 \\ -0.54})\cdot10^4$\\\hline
        \hspace{0.15cm}$6.0\cdot10^{15}$ FZ \newline\hspace{0.02cm}$6.0\cdot10^{15}$ EPI &\vspace{0.01cm}$(2.32\substack{+ 1.91\\ -1.07})\cdot10^4$ & \hspace{0.6cm}$(2.14\substack{+ 1.74\\ -0.98})\cdot10^4$ \newline $(1.20\substack{+0.85 \\-0.51})\cdot10^4$ \\\hline
          $8.0\cdot10^{15}$ FZ&  $(3.36\substack{+2.76 \\ -1.55})\cdot10^4$ &$(3.10\substack{+ 2.51\\ -1.42})\cdot10^4$\\\hline
          $1.5\cdot10^{16}$ EPI&  $(6.61\substack{+5.43 \\ -3.05})\cdot10^4$ & $ (3.41\substack{+ 2.41\\ -1.44})\cdot10^4$\\
        
         \hline
             \end{tabular}

    \label{tab:Re-calculation}
\end{table}

The reasons for the observed differences between the extracted parameters of this campaign and the standard Hamburg model can be attributed to different factors. First, the general difference of n-type and p-type sensors. In other studies in recent years a later minimum than the 90 minutes from the Hamburg model has been observed in ATLAS p-type sensors as well \cite{LDiehl_ptype}, hinting towards a generally slower annealing. 
Second, it is unknown how much the material composition, such as oxygen concentration or bulk resistivity, differs between the Hamburg model samples and the current campaign. It was already shown that oxygen enrichment has a significant effect on the radiation hardness of the samples \cite{ruzin2000radiation}, thus an effect on the annealing behavior is very likely. However, the exact oxygen concentrations of the used samples are not disclosed by the manufacturer, so a direct comparison is not possible.  
Third, the fluences considered in this campaign are at least a magnitude higher than in the dataset of the Hamburg model. It is well known that at high fluences effects not significant at lower fluences start becoming important, such as the double junction effect or the presence of a constant low field through the entire sensor volume already at low voltages, contradicting the standard concept of a depletion voltage \cite{VERBITSKAYA2006528,Agram:23731}. This has an effect on the sensors' performance, but it is not known if this impact changes during the annealing. It is unclear if the Hamburg model description is suitable for these high fluences above $1\cdot10^{15}\rm n_{eq}/cm^{2}$, since the model for the leakage current is not applicable to the current data set as well. Additionally, it is not clear whether the low-resistivity handling wafer of the epitaxial sensors starts to become active at high fluences. 

\begin{figure}
    \centering
    \begin{subfigure}{0.47\textwidth}
        \includegraphics[width=\textwidth]{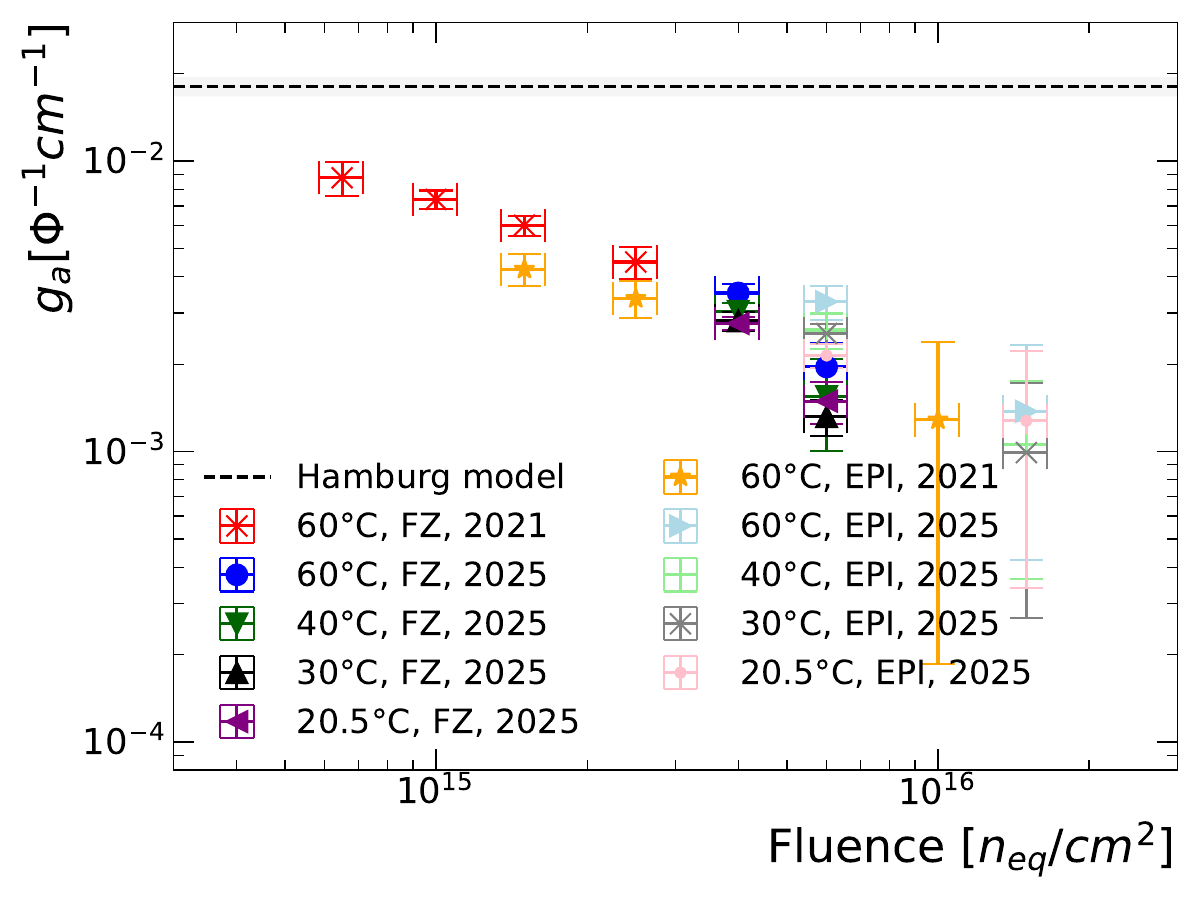}
        \caption{Beneficial annealing: $g_a$}
        \label{fig:g_a}
    \end{subfigure}
    \begin{subfigure}{0.47\textwidth}
        \includegraphics[width=\textwidth]{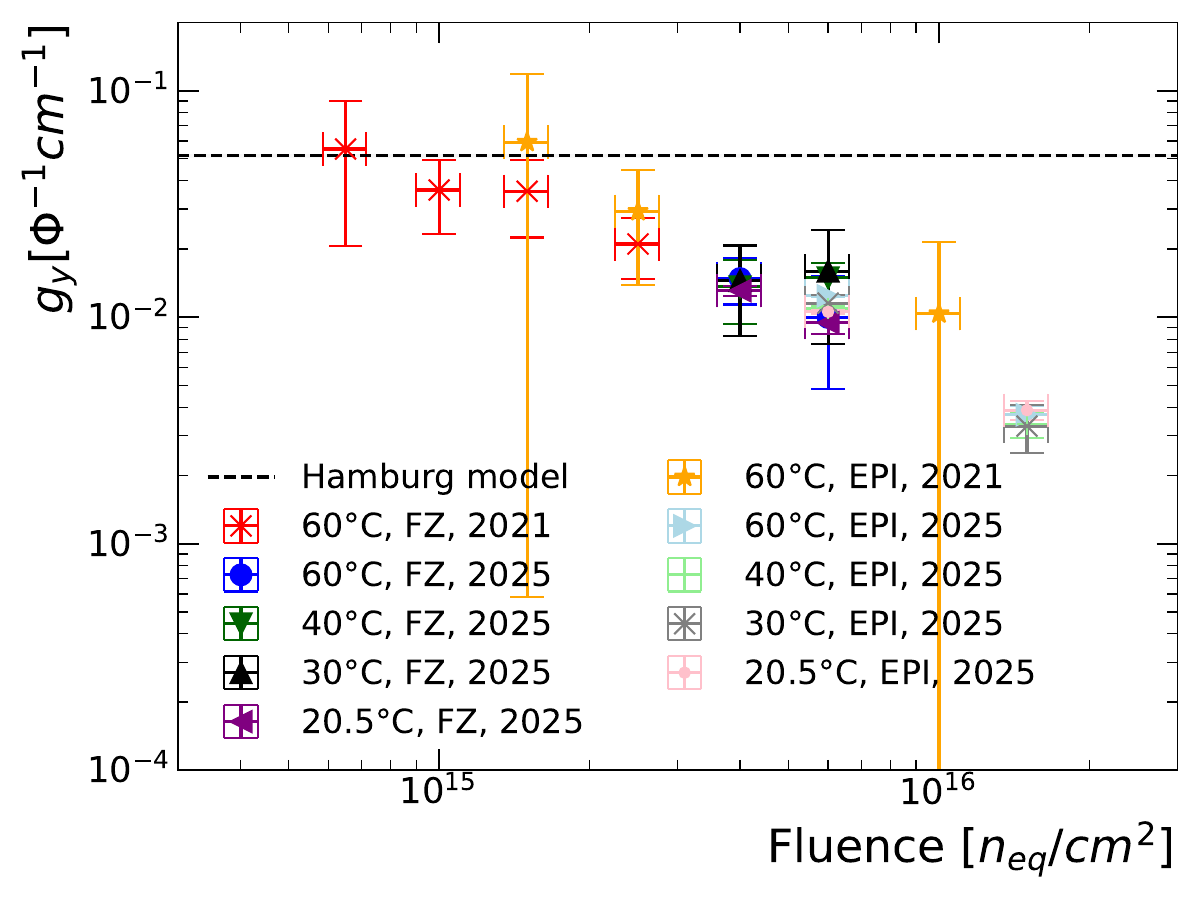}
        \caption{Reverse annealing: $g_\gamma$}
        \label{fig:g_y}
    \end{subfigure}
        \begin{subfigure}{0.47\textwidth}
        \includegraphics[width=\textwidth]{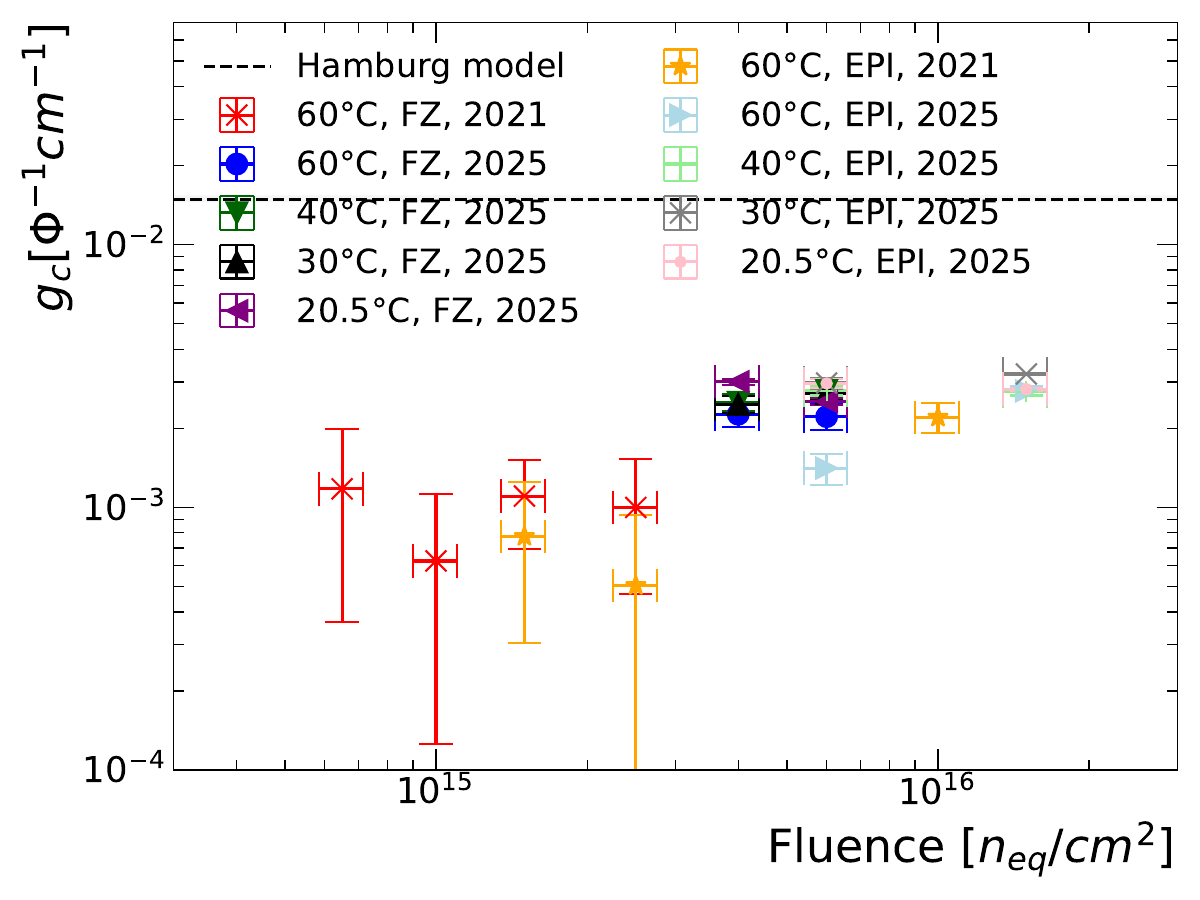}
        \caption{Stable damage: $g_c$}
        \label{fig:g_c}
    \end{subfigure}
\caption{Defect introduction rates in dependence of the fluence, separated for the two material types, annealing temperatures and measurement campaigns. The values of the 21 Campaign are taken from \cite{Kieseler_2023}.}
\label{Fig:gs}
\end{figure}

\paragraph{\textbf{Defect introduction rates}}

The last parameters to evaluate are the defect introduction rates $g_a, g_\gamma$ and $g_c$ from Equations \ref{eq:gc}--\ref{eq:g_y}, describing the beneficial annealing, the reverse annealing, and the stable damage.
The extracted parameters are presented in Figure~\ref{Fig:gs} in dependence of fluence, alongside of values extracted from a previous annealing campaign of identical samples in 2021 at $60^\circ$C to extend the fluence range \cite{Kieseler_2023} as well as the values expected for high resistivity material from the Hamburg model.

For both the beneficial and reverse annealing introduction rates $g_a$ and $g_\gamma$  a dependence on the fluence is observed, while there is no dependence on the annealing temperature as expected. 
The extracted parameters $g_a$ and $g_\gamma$ are in the right order of magnitude, and for the reverse annealing they are in agreement with the Hamburg model value for lower fluences. 
The introduction rate for the stable damage $g_c$ is measured to be one magnitude smaller than expected from the Hamburg model, while no strong fluence dependence is visible, only a small shift to higher values for fluences above $3\cdot 10^{15}\,\rm n_{eq}/cm^{2}$. The difference in stable damage between this campaign and the Hamburg model could be a further pointer toward a difference regarding the material properties and its radiation hardness. 
Contrary to the annealing time constants and the timing of the minima, there is no clear difference visible between floatzone and epitaxial material in the defect introduction rates. 

It has to be mentioned that the saturation voltage is influenced by the chosen measurement temperature and frequency. Hence, the extracted effective doping concentrations are not reliable regarding their absolute values. While this does not affect the annealing parameters $g_a$ and $g_\gamma$ significantly, it could have a larger effect on the extracted parameters for the stable damage $g_c$, as there the absolute values of the effective doping concentration have a stronger impact. Apart from the measurement method, there are several other factors that could contribute to the observed differences in $g_c$. Other studies revealed different $g_c$-values for different thicknesses in epitaxial diodes \cite{lindstrom2006radiation}, while a similar study showed differences in the donor generation rates between neutron and proton irradiation \cite{lindstrom2006epitaxial} as well as different thicknesses. The donor generation rates have an influence on the stable damage defect introduction rates. Therefore, the thickness of the sensors as well as the material could have an influence as well. Generally, most of the compared values of $g_c$ have been measured on n-type silicon sensors. In other studies \cite{mandic2025unusual, LDiehl_ptype} it has been observed that the general annealing behavior of p-type diodes varies from the expectations of the n-type based Hamburg model, which could result in deviations for the $g_c$-values as well. In another study it was shown that neutron irradiated epitaxial sensors have a lower $g_c$-value than proton irradiated samples, while p-type samples have lower $g_c$-values than n-type samples for both irradiation types \cite{kaska2010study}. The measured value for a $150\,\upmu$m thick neutron irradiated p-type epitaxial sensor was $(3.7\pm0.2)\cdot10^{-3}\rm{cm}^{-1}$ \cite{kaska2010study}, which is similar to the values measured in this study. 

A low-fluence campaign is currently ongoing to investigate if the observed fluence dependence could be correlated to the increase of influence of high-fluence effects. Thus, extending the dataset will be beneficial to evaluate if the deviations from the Hamburg model are related stronger to a difference in material, or to a potential breakdown of the model at fluences above $2\cdot10^{15}\rm n_{eq}/cm^{2}$.

\section{Summary and Conclusion}
In this study the annealing behavior at different temperatures of p-type silicon sensors irradiated to fluences between $2\cdot10^{15} -1.5\cdot10^{16}\rm 1 MeV n_{eq}/cm^{2}$ has been evaluated using leakage current, capacitance and charge collection measurements. Following the Hamburg model description ~\cite{Moll:2018fol}, the key annealing parameters have been extracted from the CV measurement results.

The general annealing behavior follows the expectations - a constant decrease of the leakage current, an increase of charge collection and a decrease of the effective doping concentration for the beneficial annealing dominated time, and vice versa for the reverse annealing dominated time.
The known phenomenon of charge multiplication due to impact ionisation after long annealing times has been observed for the sensors annealed at $60^\circ$C for which the annealing has progressed furthest into the regime dominated by reverse annealing. 

While the general course of the leakage current followed the expectations, the model used to describe and fit the data used by the Hamburg model was found not to be applicable for the datasets. It is under investigation whether this is a general change in behavior, an effect of the high fluences or affected by the lack of data points recorded in the short term annealing time range. 

Applying the Hamburg model fits to the effective doping concentration for different annealing temperatures exhibited two main differences from the expectations. First, the annealing time constants are generally higher than expected, resulting in a generally slower annealing and a later minimum. Secondly, the temperature dependence changed especially for the reverse annealing. The lower activation energy results in faster annealing at lower temperatures than expected compared to the standard $60^\circ$ C, which is done to simulate the annealing occurring during shutdown periods in the actual detectors. 

Additionally, differences were observed between the floatzone and the epitaxial material. Their origin is not understood yet, whether it is connected to high fluences, the handling wafer, oxygen or impurity concentration or the sensor thickness. 
Some hints toward a fluence dependence have been observed, e.g., in the defect introduction rates or the annealing time necessary to reach the minimum in effective doping concentration. 

The deviations from the Hamburg model could be influenced by several different parameters, namely, possible differences between p-type and n-type sensors, higher fluence range to which the sensors were irradiated, and oxygen concentration.
In order to investigate these open questions, a second long-term annealing campaign has been started, using fluences between $1\cdot10^{13}\rm n_{eq}/cm^{2}$ to $\rm2\cdot10^{14} n_{eq}/cm^{2}$, comparable to the original dataset used to establish the Hamburg model, and an increased number of samples, covering different thickness and materials per fluence to assess potential thickness, material and fluence dependences. The lower fluences will allow an extraction of the saturation voltage and thus effective doping concentration  for all thicknesses through the TCT-measurements, independent of frequency and temperature and thus will grant a better access to the absolute values of the effective doping concentration. Initial results of this low-fluence campaign are expected in Summer 2026. Additionally, a campaign investigating the effects of successive irradiation and annealing is ongoing, with results expected in the upcoming months. This campaign mimics CMS' operating scenario at the HL-LHC, with a fraction of the end-of-life fluence accumulated in Run 4, annealing in the Long Shutdown 4, and remaining fluence accumulated in the final Run 5, by doing a two-step irradiation with in-between annealing.  

\acknowledgments

We thank  Ruddy Costanzi, Matthias Kettner and David Walter for help in upgrading the Particulars setup. Furthermore, we would like to thank the CERN EP-DT SSD group and in particular Michael Moll for their support and helpful discussions, as well as the HGCAL editorial board for their thorough reviews.  We also would like to thank Marcello Mannelli for proposing the program of the study and proposing the development of the infrastructure for simultaneous measurement of the IV, CV and CC.
Moreover, we thank the CERN EP R\&D programme for funding the Particulars setup and the CMS HGCAL silicon sensor group for the discussions about the results presented here.
This work has been supported by the Wolfgang Gentner Programme of the German Federal Ministry of Education and Research (grant no. 13E18CHA) and the Alexander-von-Humboldt-Stiftung.
\newpage
\bibliographystyle{JHEP}
\bibliography{mybibfile.bib}

\end{document}